\documentclass[natbib209,12pt,preprint]{aastex}

\newcommand{\msun}{M$_\odot$}
\newcommand{\rsun}{R$_\odot$}
\newcommand{\kms}{km~s$^{-1}$}

\shorttitle{Galactic PNe from single stars and binaries}
\shortauthors{Moe \& De Marco}

\begin{document}

%\title{Planetary nebula population synthesis. I The number of galactic planetary nebulae produced by single stars}
\title{Do most planetary nebulae derive from binaries? I Population synthesis model of the galactic planetary nebula population produced by single stars and binaries}

\author{Maxwell Moe\altaffilmark{1,2} \& Orsola De Marco\altaffilmark{3}}

\altaffiltext{1}{Department of Astrophysical and Planetary Sciences and CASA, University of Colorado,\\
389-UCB, Boulder, CO 80309}
\altaffiltext{2}{REU student at the Astrophysics Department of the American Museum of Natural History}

\altaffiltext{3}{Astrophysics Department, American Museum of Natural History, \\
Central Park West at 79$^{th}$ Street, New York, NY 10024 }

\begin{abstract}
We present a population synthesis calculation to derive the total number of planetary
nebulae (PN) in the Galaxy that descend from single stars and stars in binary systems.
Using
the most recent literature results on galactic and stellar formation as
well
as stellar evolution, we predict
the total number of galactic PNe with radii $<$0.9~pc
to be ($4.6\pm1.3) \times 10^4$.  We do not claim this to be the complete population, since there can be visible 
PNe with radii larger than this limit. However, by taking this limit, we make our predicted population 
inherently comparable to the observationally-based value of Peimbert, who determined ($7200\pm1800$)~PNe 
should reside in the Galaxy today. Our prediction is discrepant with the observations at the 2.9~$\sigma$ level, 
a disagreement which we argue is meaningful in view of our specific treatment of the uncertainty. 
We conclude that it is likely that only a subset of the stars thought to be capable of making a visible PN, actually do.  
In the second paper in this series, an argument will be presented that the bulk of the galactic PN population
might be better explained if only binaries produce PNe.

The predicted PN formation rate density from single stars and binaries is ($1.1 \pm
0.5$)$\times10^{-12}$~PN~yr$^{-1}$~pc$^{-3}$ in the local neighborhood.  
This number is lower than the most recent PN birthrate density estimates (2.1
$\times10^{-12}$~PN~yr$^{-1}$~pc$^{-3}$), which are
based on local PN counts and the PN distance scale, but more in line with the white dwarf 
birthrate densities determined by Liebert et al. (($1.0 \pm
0.25$)$\times10^{-12}$~WD~yr$^{-1}$~pc$^{-3}$). The predicted PN birthrate density will 
be revised down, if we assume that only binaries make PNe. This revision will imply that the 
PN distance scale has to be revised to larger values.

\end{abstract}


\section{Introduction}
\label{sec:introduction}
%AGB evolution produces post-AGB stars with or without a PN, which then become WDs. WDs are,
%however, also produced by post RGB star that never ascended the AGB.

%PN birthrates should be smaller than WD birthrates since they do not account for all PN, but they
%are not. There are flaws in the way PN birthrates are calculated.

The origin of planetary nebulae (PNe) has been the subject of a hot debate for the last thirty years
(for a list of references see \citealt{Soker1997}).
The bone of contention has been the mechanism by which the large majority of PNe achieve axi- and
point-symmetric shapes. The proposed mechanisms can be divided into
two classes. The first class of mechanisms 
involves single stars, proposing a combination of magnetic fields and/or stellar rotation to achieve an
axi-symmetric mass-loss during the Asymptotic Giant Branch (AGB) phase \citep{GarciaSegura2004, Matt2004,Blackman2004} 
into which the fast central star wind expands. The second class of mechanisms proposes that
PNe with non spherical morphologies must have suffered an encounter with a binary companion (whether a close encounter such as
a common envelope or a lesser interaction such as gravitational focussing). Depending on the binary separation,
mass ratio and other parameters the PN ejecta take on different shapes.

A simple test of which of the two classes of models is more likely to be correct is
to determine the PN central star binary fraction.
About 10--15\% of all PN are known to harbour close binary central stars \citep{Bond2000}. 
These were discovered with photometric surveys that rely on heating of one hemisphere 
of the companion by the primary, or by ellipsoidal variations. 
Since heating effects and ellipsoidal variations quickly diminish with binary 
separation, only binaries with periods smaller than a few days can be detected.  
Surveys aimed at detecting visual binaries have been carried out \citep{Ciardullo1999} 
and, once again, about 10\% of all observed central stars appear to have a distant 
companion (separations $>>$ 100~AU). The companions of the latter group are unlikely to have played any significant role in shaping the PN, although it is not excluded that these wide binaries might harbor a third companion orbiting closer to the primary.

It is clear that until surveys are carried out that can systematically detect binaries with 
periods between a few days and a few years, we cannot know the actual PN binary fraction. 
Radial velocity surveys are ideal to fill the period gap, but their execution is 
particularly tricky, since relatively long observing runs are needed on moderate aperture 
telescopes with intermediate-to-high resolution spectrographs. \citet{Mendez1989} reported on the results of 
a radial velocity survey of 28 central stars 
carried out at intermediate resolution. However, he only carried out one or two measurements per star,
severely limiting the discovery potential of his survey (he had no conclusive evidence to prove the binarity of
any stars in his sample) and precluding the determination of periods. 
\citet{Sorensen2004} carried out a new survey of 33 central stars, at lower resolution, 
but where many measurements were taken for every star. 
They report about 40\% radial velocity variabilty in their sample of 33 central stars, 
although no periods could be determined. 

More recently,
\citet{DeMarco2004} and \citet{Afsar2005} carried out radial velocity surveys with resolution comparable to the Mendez (1989) and Sorensen \& Pollacco (2004) studies,
respectively. Their samples (11 and 19 objects, respectively)
provided tantalizing evidence
that a much larger fraction, possibly as high as 90\%,
of central stars might have companions. 
Periods were not detected by these surveys, leaving the possibility that in some cases the radial velocity variability might be due to stellar wind variability. From additional observations (De Marco et al., in preparation, \citealt{Bond2005})  at echelle resolutions  periods between 4 and 5 days were determined for both stars observed, which where drawn from the sample of De Marco et al. (2004; 
IC4593 and BD+33~2642). This fuels the empirically-based suggestion that  the PN central star close binary fraction might indeed be very high.

Radial velocity surveys of central stars of PN are, however,  far from being complete so that an observational
answer to the question of the exact PN central star binary fraction is not yet in hand.
We have therefore carried out a population synthesis calculation
to revisit the issue of how many PNe are predicted in the Galaxy today.  The {\it empirical} suggestion that 
most-to-all PNe are in short-period binary systems, should be supported, or at least consistent, 
with current theories of stellar evolution, galactic history,
star formation and knowledge of related stellar populations (such as AGB stars and white dwarfs [WDs]).
This suggests that  population synthesis calculations which assume that single stars {\it do} make PNe should predict an overabundance of PNe compared to observations. 
In this paper (the first in a series), we address single star evolution and the total PN population.
We analyze all the sources of error and state the issues involved with PN and WD counts and
birth rates. In the second paper of this series (De Marco et al., in preparation, hereafter Paper~II),
we will address the issue of binarity as the main evolutionary channel for the production of PNe.  

In \S~\ref{sec:BotE} we outline the phases of our population synthesis calculation as if it were
a back of the envelope calculation. In \S~\ref{sec:popsyn} we give a detailed explanation of our model method,
as well as of our assumptions concerning the initial mass function (IMF; \S~\ref{ssec:IMF}), 
stellar lifetimes (\S~\ref{ssec:lifetimes}), the mass of the Galaxy (\S~\ref{ssec:massMW}), the
star formation history (SFH; \S~\ref{ssec:SFH}), the age-metallicity relation (AMR; \S~\ref{ssec:metallicities}) 
and the PN visibility timescales (\S~\ref{ssec:PNLifetimes}). In \S~\ref{sec:results}, we present our results, we explain the
uncertainties, and carry out a detailed comparison with observations and past
results from the literature, setting the stage for the binary synthesis (Paper~II). 
In \S~\ref{sec:nrPNbulgeGCs} we compare our PN population predictions with the observed populations of the galactic
bulge and galactic globular cluster (GC) system. We conclude in \S~\ref{sec:conclusion}.

\section{The back-of-the-envelope calculation}
\label{sec:BotE}

The genesis of this model is the back of the envelope
calculation of \citet{DeMarco2005}, which we outline here because it provides an overview of the process.

In order to predict the total number of PNe that live in the Galaxy today, 
\citet{DeMarco2005} started with
the total number of stars in the Galaxy, determined by normalizing the total galactic luminous mass
($7.5 \times 10^{10}$~\msun ; \citealp{Dehnen1998,Mera1998}) to the average mass of a star (0.54~\msun),
calculated using the IMF \citep{Kroupa1993,Chabrier2003big}. 
This resulted in $1.4 \times 10^{11}$ stars, where single stars count as one, and binary stars count as two.
60\% of the 
observed G- and late F-type main sequence stars are found to be in binaries \citep{Duquennoy1991}. Assuming that as the general stellar binary fraction, 
the total number of stellar
systems (where by {\it stellar system} we mean a single star or a binary systems, i.e., we are counting
binaries as {\it one} system) is $8.8 \times 10^{10}$. Of these, we considered only
single stars and primary stars in binaries with masses large enough to have evolved off the main
sequence in the age of the Galaxy. To do this, we considered the Galaxy as a coeval population with an age of
8.5~Gyr \citep{Liu2000,Zoccali2003}. The turn-off mass of such a population is 1.05~\msun\ (based on the stellar lifetime of
\citet{Portinari1998}). Using the IMF,
we calculated that the total number of galactic stars between 1.05~\msun and 10~\msun (the upper mass
limit above which a star explodes as a supernova and does not make a PN; \citealt{Iben1995}), is 8.4\% of all stars. This
results in $7.39 \times 10^9$ systems. The
mean mass of the population between 1.05 and 10~\msun\ is 1.78~\msun\ and the lifetime associated with such a
stellar mass is $1.3 \times 10^9$~yr \citep{Bressan1993,Schaller1992}. Finally, we adopted 20\,000~yr as the mean PN
visibility time. The number of PNe in the Galaxy was then estimated by multiplying the number of stars in the 
prescribed mass range, by the ratio of PN visibility time to mean stellar
lifetime ($7.39 \times 10^{9} \times 20\,000 / 1.3 \times 10^9$ = 113\,700 PNe).

Even accounting for a substantial uncertainty, this number is much higher than even the most optimistic observational estimates of the total number of
galactic PNe. The exact number of PN in the Galaxy is hard to establish due to the difficulty of
observing along the galactic plane. \citet{Parker2003} have counted 2000-2500 galactic PNe in the most
extensive survey to date, but this is bound to be a lower limit because of extinction on the galactic plane would obscure
a fair number of distant PNe.
In \S~\ref{sec:results} 
we will go into greater detail of the total number of galactic PNe determined via observationally-based methods. 
For now, suffice it to say that our preferred observationally-based  galactic PN population size is 7200$\pm$1800 
\citep{Peimbert1990,Peimbert1993}. 

The discrepancy between the theoretically-predicted and observationally-derived galactic PN population size, prompted us to refine our prediction of the number of galactic PNe, by using a more sophisticated method. As we will see, even after refining the calculation, the total number of PN predicted in the Galaxy is still too high compared to observations and leads one to consider the sources of such a discrepancy.

\section{The Population Synthesis Method}
\label{sec:popsyn}

We have constructed a stellar population synthesis code to enable us to predict
the number of PNe in the Galaxy today. Although the method is analogous to the 
back-of-the-envelope calculation described above, all assumptions have been refined,
and the calculation is carried out using time bins, instead of averages.

The main refinements can be summarized as follows: we have divided the Galaxy into its four main components, namely
the spheroid, the bulge, the thick and thin disks. (Note that we never use the word {\it halo}, to avoid confusion with
the dark matter halo. We use instead the term {\it spheroid} to mean
the luminous halo including the globular clusters [GCs]). For each of these four components we use different mean ages
as well as different IMFs, star formation rates (SFR) and metallicities. We also adopt an AMR
when determining
metallicity-dependent stellar lifetimes and PN visibility times.
We calculate the PN visibility time using
nebular kinematics and stellar evolution arguments. We also account for the fact that some red giant branch (RGB)
stars undergo common envelope interactions, which reduce their envelope mass below the the critical value
needed to ascend the AGB. These stars are unlikely to ascend the AGB and make a PN and are therefore not counted.

The SFH of the Galaxy is split into 649 
time bins. The time variable, $t$, starts at $t_0=0$~Gyr. Time bin boundaries are then determined by the following equation:

\begin{equation}
t_{i+1} = (0.1~{\rm Gyr}) \times 0.9923607^{i} + \sum_{j=0}^{i} t_{j}
\end{equation}

so that  $t_{(650)}=13$~Gyr (adopted as the age of the oldest galactic component
- see below).  In this way, the
resolution near time 0~Gyr is 0.1~Gyr, but near $t=13$~Gyr it is less than 1~Myr. This level of refinement avoids large
rounding errors produced by the steep gradient of stellar lifetimes, since stars more massive than 2.0~$M_{\odot}$
have evolved off the main sequence in the past giga-year, but stars more massive than 4.0~$M_{\odot}$ have turned off
only in the past 100~Myr. 

Consider bin boundaries at $t_i$ and $t_{i+1}$.  The corresponding turn off masses
$m_i$ and $m_{i+1}$ at the bin boundaries
are calculated by determining the masses of stars that have stellar lifetimes 
of (13~Gyr - $t_i$) and (13~Gyr - $t_{i+1}$), respectively (note that the stellar lifetimes are computed according to
metallicity-dependent stellar evolutionary tracks and the AMR; \S~\ref{ssec:metallicities}). 
The relative number of stars produced
in this time interval is calculated from the SFR function and this value is multiplied by the fraction of stars
between $m_i$ and $m_{i+1}$ determined from the IMF ($\phi(m)$; \S~\ref{ssec:IMF}). This results in the
total number of PNe being produced today whose progenitors formed during that time interval. The fraction
of PNe currently detectable is given by the ratio of the PN visibility time
to the width of the time bin ($t_i-t_{i+1}$).  
Summing all of the bins' contributions from 0 to 13~Gyr
and all components of the Galaxy, gives the total number of PNe currently detectable in the Galaxy.

Below we give details of each phase of the calculation, including an accounting of the
uncertainty.

\subsection{The Initial Mass Function}
\label{ssec:IMF}

The IMF represents the mass distribution of
stars at the time of their formation.  We use the IMF                     
to determine the fraction of stars in a population
that are massive enough to evolve off the main sequence during the age
of that population but less massive than $8.0~M_{\odot}$ (in \S~\ref{sec:BotE} we used 10~\msun\ as an upper limit,
but the exact value is not very important as there are relatively few stars with high mass), an upper mass limit above which stars
tend to become supernovae rather than making PNe \citep{Iben1995}.
We will also use the IMF to scale the total luminous mass of the Milky Way and determine the total number of
galactic stars.

\citet{Salpeter1955}
first proposed
that the IMF consists of a single power law over most of the
stellar mass range ($0.4 M_{\odot} < m < 10 M_{\odot} $): 

\begin{equation}
\xi (m) = k \> m^{-x},
\end{equation}

\noindent where $\xi (m) dm$ is the {\it mass} of stars in the interval $m$ to $m+dm$,
$x$ is the power law index (determined by \citet{Salpeter1955} to be 1.35)
and $k$ is a normalization constant.
Other authors (e.g., \citealp{Kroupa1991, Kroupa1993}) represent the
IMF as a distribution {\it by number} instead of by mass:

\begin{equation}
\phi (m) = k^\prime \> m^{-\alpha},
\end{equation}

\noindent where the total {\it number} of stars in the interval $m$ to $m + dm$ is
$\phi (m) dm$, $k^\prime$ is a different constant, and
$\alpha = x + 1$, since:

\begin{equation}
\int_{m_{low}}^{m_{up}} m \> \phi(m) \> dm=\int_{m_{low}}^{m_{up}} \xi (m) \> dm,
\end{equation}

\noindent where $m_{low}$ and $m_{up}$ are the lower and upper mass limits under
consideration.

Unresolved binary stars were taken into account by \citet{Kroupa1991, Kroupa1993},
who developed a three-segment power-law IMF so that random, uncorrelated
pairings of stars produced the observed (e.g., \citealt{Duquennoy1991, Fischer1992, Eggleton1989} and \citealt{Tout1991})
frequencies and mass-ratio distributions of binaries.
\citet{Kroupa2001} showed that for the average-age galactic disk, the 
IMF exponent changes:

\begin{eqnarray}\label{alpha.eqn}
\alpha_0 & = & 0.3, \; \; 0.01 \leq m <0.08,\nonumber \\
\alpha_1 & = & 1.3, \; \; 0.08 \leq m <0.5,\\
\alpha_2 & = & 2.3, \; \; 0.5 \leq m. \nonumber
\end{eqnarray}

The decrease in the slope and eventual turnover in the
sub-stellar mass regime has led some authors to adopt a log-normal mathematical
representation of the IMF of stellar {\it systems} (i.e., taking into account binaries; 
\citealp{Chabrier2003big, Chabrier2003small}):

\begin{equation}
\xi (m) = k
\cases{A \> exp \> [\frac{-(log \> m - log \> m_c)^2}{2\sigma^2}] & $ \; m \leq m_{norm}$,\cr
 B \> m^{-1.3} & $ \; m > m_{norm}$,\cr}
\end{equation}

\noindent where the $A$ and $B$ coefficients are used to ensure continuity at the 
segment transitions. For the average age disk, $m_{norm} = 1.0$,
$m_c = 0.22$ and $\sigma = 0.57$, while  
for the spheroid $m_{norm} = 0.7$,
$m_c = 0.22$, and $\sigma = 0.33$.

In addition, evidence is accumulating that the IMF
may be time dependent, with
earlier epochs favoring a higher abundance of more
massive stars \citep{Kroupa2001, Chabrier2003big, Lucatello2005}, so that
$\alpha$ is 0.5 higher between 0.08 $M_\odot$ and 1.0 $M_\odot$ for
the present day IMF. In this work, we use the $\alpha$ values from Eqs.~\ref{alpha.eqn},
without accounting for the time dependency. 

\begin{figure}
\centering  
\includegraphics[width=3.6in]{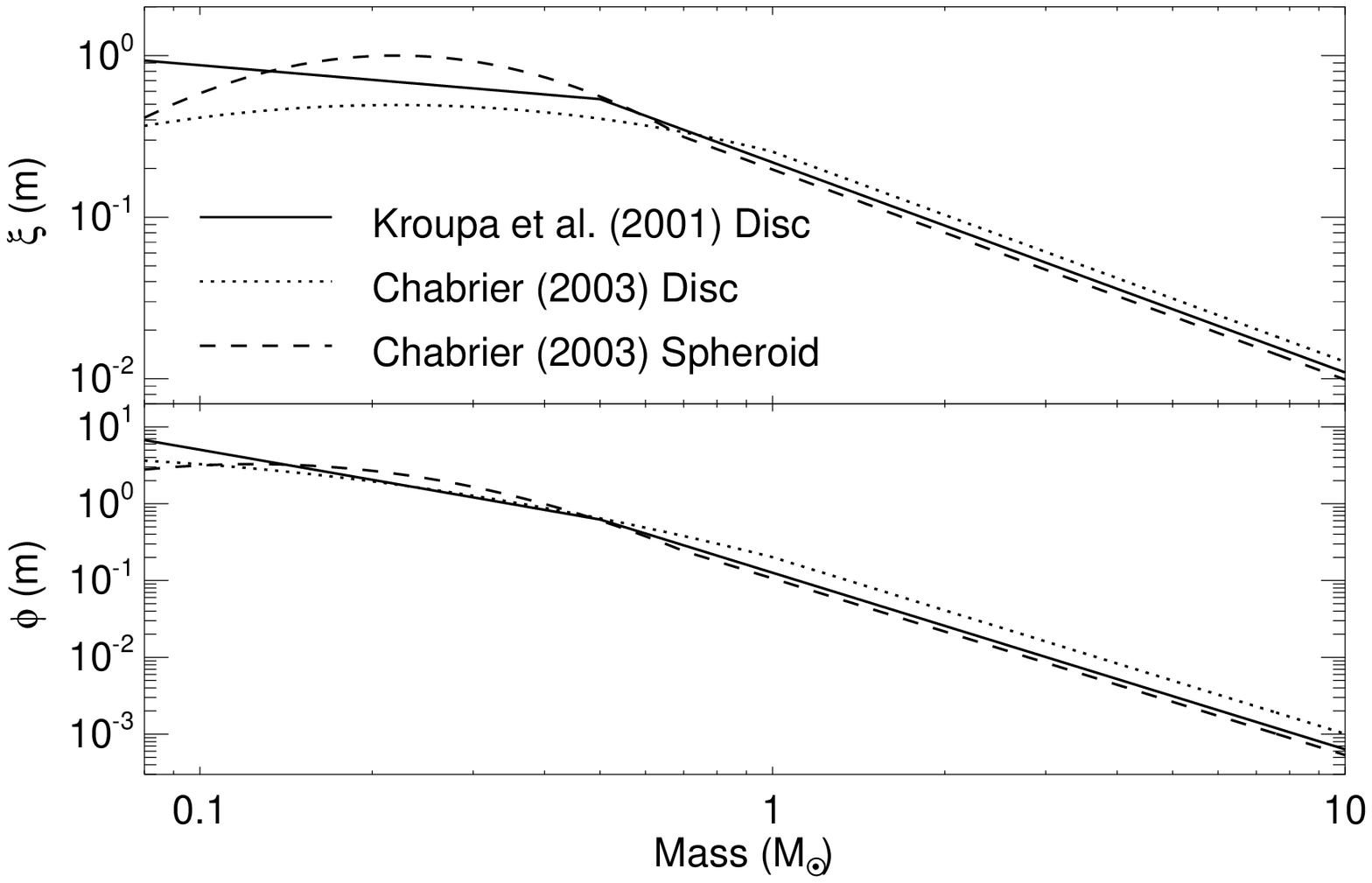}
\caption{The three IMFs (normalized) that will be considered in our analysis.  Top panel is the
relative mass of stars per mass interval while bottom panel is the number of stars
per mass interval.  }
\label{IMF.fig}
\end{figure}

Since there is still much controversy over the IMF of
the lower mass regime, as well as whether the IMF is time
and environment dependent, we will consider three IMFs: the 
average-age disk IMF for the entire Galaxy of \citet{Kroupa2001}, that of 
\citet{Chabrier2003small}
and, finally, the spheroid IMF of \citet{Chabrier2003big, Chabrier2003small} which we will
apply to the spheroid, bulge and thick disk.  These
three IMFs are plotted in Fig.~\ref{IMF.fig}, both by mass, $\xi(m)$, and by number,
$\phi(m)$. A comparison of the PN population when using either of the two disk IMFs will
allow us to assess the error incurred because of uncertainties in the IMF prescription, while comparing the
results using either a disk IMF for all galactic components or using the spheroid IMF of \citet{Chabrier2003big, Chabrier2003small} 
when dealing with star formation in older components
will allow us to evaluate the uncertainty incurred because of actual changes in the IMF 
as a function of time and environment.  

For our analysis, we will use the IMF between 0.08 $M_{\odot}$ and 120 $M_{\odot}$ to determine the total number of stars from the luminous matter in the Galaxy. The lower limit was set to a stellar value, since the galactic mass is determined from fitting the luminosity
of main sequence stars and their giant descendants (see \S~\ref{ssec:massMW}) . 
We can then integrate the three IMFs in Fig.~\ref{IMF.fig} to obtain the median
main sequence mass for galactic stars. This median mass will be different if we consider the
IMF expressed by mass or number (upper and lower panels, respectively, in Fig.~\ref{IMF.fig}).
The median mass of a star when considering the IMF by mass is 1.30~$M_{\odot}$, 1.93~$M_{\odot}$, and 1.01~$M_{\odot}$, for the three IMFs of \citet{Kroupa2001} (disk),
\citet{Chabrier2003small} (disk), and \citet{Chabrier2003big} (spheroid). The median
mass of a star when considering the IMF by number is 
0.2~$M_{\odot}$, 0.32~$M_{\odot}$, and 0.26~\msun, for the three IMFs, respectively.
This means that
half of the {\it mass} of a coeval population of main sequence stars is locked up in stars with masses greater than
about 1.3 $M_{\odot}$ (taking the \citet{Kroupa2001} as an example), 
but the typical mass of a {\it star} in that population would be about 0.24
$M_{\odot}$.

\subsection{Stellar Lifetimes}
\label{ssec:lifetimes}

The next step in determining the number of PNe in the Galaxy, is to determine how many
galactic stars have had enough time to evolve to the PN phase. If the Galaxy were
a coeval population, only stars more massive than the main sequence turn-off mass of the population
would be available to evolve to the PN phase. Since the Galaxy is not coeval, we
have to consider its SFH (\S~\ref{ssec:SFH}). Before we do so, we want to determine the lifetime
corresponding to each mass bin, since that, together with the age of each star (determined according
to the SFH), will determine whether that star has evolved off the main sequence and into a PN.

Stellar lifetimes depend on stellar masses, but also have a weak dependence on the metallicity $Z$.
We compare the lifetimes calculated from the stellar models of
\citet[Z=0.02 \& Z=0.001]{Schaller1992} and of the Padova group:
\citet[Z=0.008]{Alongi1993}, \citet[Z=0.02]{Bressan1993}, \citet[Z=0.0004
\& Z=0.05]{Fagotto1994a}, \citet[Z=0.004 \& Z=0.008]{Fagotto1994b}, 
\citet[Z=0.1]{Fagotto1994c}, and \citet[Z=0.0001]{Girardi1996}.  These models
considered the most up to date radiative opacities at that time \citep{Rogers1992, 
Iglesias1992}, yielding higher accuracy. 
A summary of the results from the Padova group can be found in the paper by \citet{Portinari1998}.
{\it In what follows, we define the stellar lifetime, $\tau_{*}(m,Z)$, to be the sum of a star's hydrogen and core helium
burning phases.}

\begin{figure}
\centering  
\includegraphics[angle=-90,width=3.6in,angle=90]{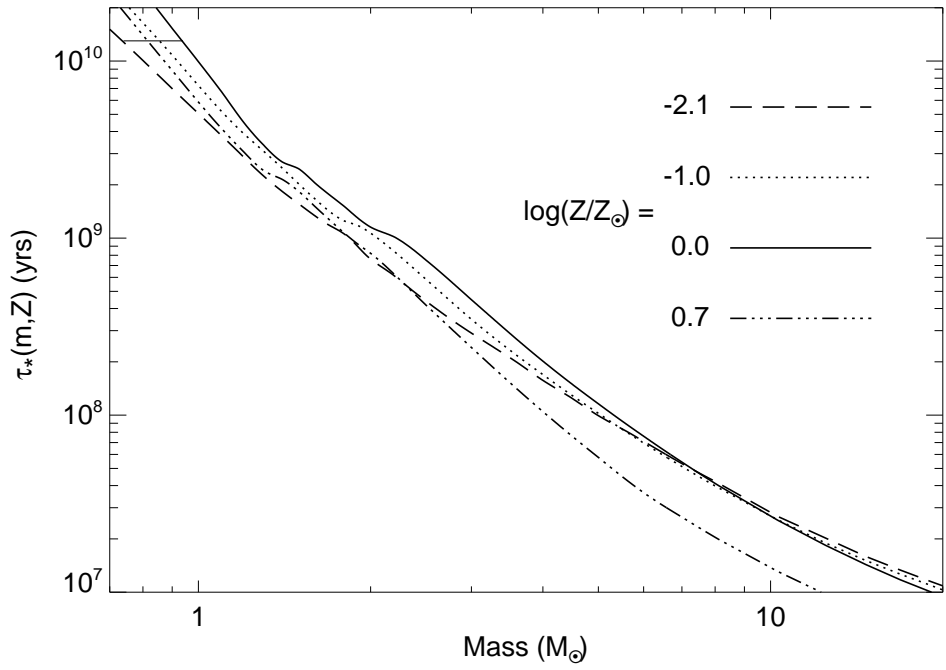}
\caption{Five isometallic lifetimes of stars as a function of mass.  The horizontal line in the
top left at 13 Gyr represents the range of masses with lifetimes equal to the age of the
universe.}
\label{Lifetime.fig}
\end{figure}

We determined a smooth function for the stellar age as a function of initial mass and
metallicity by 
cubic spline interpolation in $log(m)$ and $log(Z/Z_{\odot})$, respectively.
Four isometallic functions obtained in this way are plotted\footnote{
We represent metallicity as the logarithmic fraction of the solar value, 
$log(Z/Z_{\odot})$, where $Z_{\odot} = 0.015$ \citep{Asplund2005, Basu2004,
Bahcall2005}.} in Fig.~\ref{Lifetime.fig}. Looking at Fig.~\ref{Lifetime.fig}, we note
that the stars with the longest lifetimes tend to be those with solar metallicity.  At 
metallicities lower than solar, the
lifetimes increase with increasing metallicity, due to higher opacities, while the
lifetimes of stars with metallicities greater than the Sun are limited by their supply of
hydrogen fuel. Thus, even
though the average metallicity of the Galaxy may be near solar metallicity, the average
lifetimes of stars in the Galaxy are less than the solar lifetimes.
If a star has a metallicity 
$log(Z/Z_{\odot})<-2.2$ or $log(Z/Z_{\odot})>0.8$, we compute its lifetime 
as if it had the boundary metallicity.

For our analysis, we assume the galactic spheroid began forming 13 Gyr ago
(see \S~\ref{sssec:SFHSpheroidThickDisk}), and thus we set
our time variable $t$ = 13.0 Gyr - $\tau_*$.
The turnoff mass for a 13 Gyr population ranges from
0.74 to 0.94~$M_{\odot}$, depending on metallicity. 

The fraction of stars between
0.74 $M_{\odot}$ and 8.0 $M_{\odot}$ is
13.7\%, 21.3\% and 11.6\%, for the \citet{Kroupa2001}, \citet{Chabrier2003small} and
\citet{Chabrier2003big} IMFs, respectively, where we use the IMF by number, $\phi(m)$, while between
0.94 $M_{\odot}$ and 8.0 $M_{\odot}$, it is 9.9\%, 15.7\% and 8.3\%, respectively.
We therefore see that if we {\it do not}
consider metallicity in determining stellar lifetime, we would incur a
substantial error  in our determination of
the number of stars available to produce today's PN population.

\begin{figure}
\centering  
\includegraphics[width=3.6in]{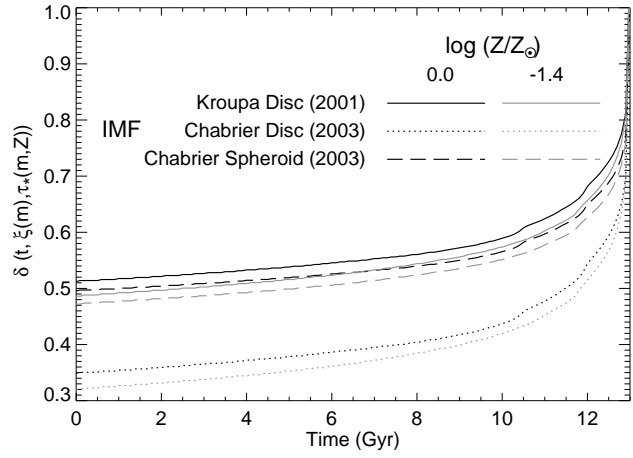}
\caption{The mass fraction of stars $\delta$ that were formed at time $t$ and
remain luminous today for varying IMFs and metallicities.}
\label{LifetimeIMFfactor.fig}
\end{figure}

The mass fraction that remains {\it luminous} (i.e., from stars that are still burning hydrogen or helium),
for a population of stars that
formed at time $t$, is quantified by:

\begin{equation}
\label{delta.eqn}
\delta(t,\xi(m),\tau_{*}(Z)) = \frac{\int_{0.08}^{m_{t}(\tau_{*}(Z))} \xi(m) \> dm}{\int_{0.08}^{120} \xi(m) \> dm},
\end{equation}

\noindent where $m_{t}(\tau_{*}(Z))$ is the mass of a star with lifetime $\tau_{*} = 13-t$, i.e. it
is the turnoff mass for the population. 
We call $\delta$ the {\it stellar evolution correction factor} and we plot it 
in Fig.~\ref{LifetimeIMFfactor.fig} as a function of time
for three IMFs and two different metallicities.
Note that if a population of stars was born as little as 1 Gyr
ago at $t$ = 12 Gyr, about 40\% of that population's mass has been converted to stellar remnants.
Also, only 30-50\% of the original mass of a population of stars that formed at the
beginning of the Galaxy remains luminous today.

\subsection{The Mass of the Milky Way}
\label{ssec:massMW}

In order to determine the number of galactic stars that are available to
produce the PN population (i.e., those between $\sim$0.8 $M_{\odot}$ and
8.0 $M_{\odot}$), we use estimates of the 
luminous matter in the Milky Way and use the IMF to determine the number of objects.
In what follows, we treat the
bulge, thin disk, thick disk and spheroid separately.  

\subsubsection{The mass of the bulge}
\label{sssec:massBulge}

\citet{Kent1992} was among the first to calculate an accurate model for the bulge, deducing that it has
an axisymmetric shape.
From the resulting density and luminosity distributions, the bulge was estimated to have a mass of
$~1.8 \times 10^{10} M_{\odot}$.  Other models incorporating gas
kinematics \citep{Binney1991}, metallicity gradients \citep{Zhao1994},
and micro-lensing events
\citep{Paczynski1991,Zhao1995, Han1995},
indicate the bulge to have a triaxial symmetry.
Luminosity fits and dynamical models of a triaxial bulge have yielded
masses ranging from $1.3 \times 10^{10}~M_{\odot}$ to $2.8 \times 10^{10}~M_{\odot}$ 
\citep{Dwek1995, Blum1995, Zhao1996, Fux1997, Fux1999,
Weiner1999, Sevenster1999a}, depending on the inclination of
the major axis to the line of sight. 

Recent deep near-infrared
surveys have fixed the bulge orientation at 10$^o$ - 20$^o$  \citep{Lopez1999, Lopez2000,
Lopez2005, Picaud2004}.
With these new constraints, the mass of the
outer bulge (r $> \> \approx$150 pc) was calculated to be ($2.3 \pm 0.4) \times 10^{10}$~M$_{\odot}$,
with about a quarter of this mass contributed by stellar remnants
\citep[and references therein]{Robin2003,Picaud2004}. Thus, for the
present day {\it luminous} mass of the bulge we adopt
($2.0 \pm 0.4) \times 10^{10}$~\msun.  This
figure accounts for the luminous mass ($1.7 \times 10^{10}$~\msun=$0.75 \times 2.3 \times 10^{10}$~\msun) 
in the outer bulge, plus an additional
$0.3 \times 10^{10}$~\msun\ contained in the high density regions of the inner bulge \citep{Lindqvist1992}.

\subsubsection{The mass of the disk}
\label{sssec:massDisk}

The mass of the disk is difficult to determine since it depends heavily
on the observed stellar surface densities, scale height, scale length, as well as the solar position
within the Galaxy. 
Gilmore \& Reid (1983) discovered that the stellar population
1-2 kpc above and below the galactic plane is different from that closer to the mid-plane.  
These {\it thick disk} stars are
older, have generally lower metallicities and higher velocity
dispersion than the stars of the {\it thin disk}. The thick disk has been found to account
for $\sim$10\% of the total luminous stellar content of the entire disk
(Cabrera-Lavers et al. 2005).  From models of the chemical evolution of the
entire disk based on observed abundances, its mass has been estimated to be in the range
$4-5 \times 10^{10} M_{\odot}$, with a gas fraction between $\sim$10 and $\sim$18\%
\citep{Prantzos1995, Boissier1999}.  On the other hand, models based primarily
on the stellar surface densities, scale length and scale height, find the stellar mass to be
in the range $2.6-3.5 \times 10^{10} M_{\odot}$, with a slightly higher gas fraction, between
16 and 25\% \citep{Dehnen1998, Naab2006, Robin2003}.  We therefore adopt 
the total luminous mass of the disk to be $(4.0 \pm 0.7) \times 10^{10} M_{\odot}$,
where the stellar mass of the thin disk accounts for ($3.6 \pm 0.7) \times 10^{10}$~\msun,
while that of the thick disk makes up the remaining $4 \times 10^{9}$~\msun.  

The value of $~4.0 \times 10^{10} M_{\odot}$ 
was chosen as the weighted average of both the higher mass disk ($4-5 \times 10^{10} M_{\odot}$, determined
via the chemical evolution model)
and the lower mass disk ($2.6-3.5 \times 10^{10} M_{\odot}$, determined via the stellar 
surface density), based on the uncertainties provided by the cited papers.
The lowest mass estimate of $2.6\times10^{10}$~\msun\ 
by \citet{Dehnen1998} included
a 25\% gas fraction, which is higher than what appears to be the more general 
consensus (15\%). We have therefore raised this mass
estimate by a factor
$0.85/0.75$, to $3.0\times10^{10}$~\msun, to account for the 
15\% gas fraction that most authors argue in favor of.  

\subsubsection{The mass of the spheroid}
\label{sssec:massSpheroid}

The spheroid is the oldest population of the Milky Way and resides in a
spherical region extending 10--20 kpc out from the galactic center (Reid 1996). 
Counting only the main sequence stars, the mass of the spheroid stars has been
estimated to be in the range $0.3-3 \times 10^{9}$~\msun , based on assumptions about
the local densities of such objects as well as relative counts compared to other
populations \citep{Edmunds1984, Chabrier1997,Gould1998, Robin2003}.
We assume the spheroid to have $2 \times 10^9$~\msun, based on the
smaller errors of the more recent estimates. The exact number is, however, not of
fundamental importance since the contribution of PNe from the spheroid is relatively small.

\subsubsection{Errors}
\label{sssec:errors}

In order to estimate the uncertainty on the total PN population derived from
determining the galactic mass we consider three distinct mass values:
a low, medium and high mass Galaxy.
For the three sets of models we adopt the
values summarized in Table~\ref{tab:mass}.

It would be more correct to use nine, instead of three, galactic mass models,
since each of the three choices for the thin disk mass could be combined with each
of the three values for the bulge mass. This method would give more weight to the intermediate
values, rather than altering our final results. Since the total number of final models is high and since most PNe derive from the thin disc,
it was decided that the slight difference in error weighting produced by the use of 9 (rather than 3)
galactic mass models is not paramount.

\subsection{The Star Formation History}
\label{ssec:SFH}

In the back-of-the-envelope model (\S~\ref{sec:BotE}), we assumed that all galactic stars formed 8.5~Gyr ago.
Using the IMF and the stellar lifetimes as a function of mass we determined the turn-off mass of an
8.5-Gyr old population to be 1.05~\msun. We then determined
the number of stars in that population that have left the main sequence and proceeded to the PN phase. 
Since the Galaxy is not a coeval population, we need to address its star formation history and use
it to determine a progression of stellar ages, instead of using an approximate mean value.

\subsubsection{The star formation history of the spheroid and thick disk}
\label{sssec:SFHSpheroidThickDisk}

Stars in the galactic halo and the spheroid GCs (which we cumulatively refer to as 
the spheroid)
began forming about 13 Gyr ago. Star formation lasted there only 4-5 Gyr, so that the spheroid has a mean age of 11.5 Gyr \citep{Harris1996,Liu2000, Bekki2001,
Gratton2003, Angeli2005}.  The thick disk is slightly younger with star formation
beginning 1-2 Gyr after the initial formation of the spheroid and
lasting for about 5 Gyr. The thick disk has a mean age of 10 Gyr \citep{Nissen1991, Marquez1994,
Fullton1996,Bensby2004}.  

\subsubsection{The star formation history of the bulge and thin disk}
\label{sssec:SFHBulgeThinDisk}

The ages of the bulge and thin disk populations are more
difficult to estimate, since star formation has continued there for most of the galactic history.
Originally, it was believed that the bulge consisted purely of old
stars (ages $>$ 10 Gyr), but new evidence points to the existence of a bulge population of
intermediate mass stars with ages between 5 and 10 Gyr
\citep{Holtzman1993}, as well as a substantial fraction of
stars near the galactic center that have formed within the past 5 Gyr
\citep{Krabbe1995, Blum1996a, Blum1996b, Narayanan1996}.
These findings should not however tempt us to adopt a low mean age for the bulge. \citet{Ortolani1995}
estimates that the intermediate mass population contributes no more
than 10\% to the total and that the young stars are confined to the
inner $\sim$200~pc of the bulge \citep{Frogel1999}, where the older population
still dominates by number \citep{Rich1999, Rich2001}.
\citet{Feltzing2000} and \citet{Zoccali2003} also advocate an old age for the bulge of 10-11~Gyr,
arguing that younger age estimates (9~Gyr; \citealp{Gerhard1993,
Wyse1997, Sevenster1999b}) are due to contamination of the bulge population by foreground stars.
Since the issue remains unresolved, 
we adopt three distinct values for the average age of the bulge: 9, 10 and 11 Gyr.  

The difficulty in measuring the average age of the thin disk arises
from the fact that the star formation rate (SFR) varies with radial distance from the 
center. \citet{Naab2006} showed that the average age of a
star in the thin disk is 5.4~Gyr, while the average age of a star in the solar
neighborhood is 4.0~Gyr. Observations of the
scale height and age distribution of stars throughout the Galaxy
as well as dynamical models of the infall of gas onto the disk
have constrained the average age of the thin disk to the range 4-7 Gyr \citep{Twarog1980,
Haywood1997a, Haywood1997b, Rocha2000, Liu2000, Chang2002, 
Naab2006}.   
We therefore adopt three values for the average age of the
thin disk: 4.5, 5.5 and 6.5~Gyr, which is the weighted mean and uncertainty of the cited papers.

\subsubsection{Star Formation History Modeling}
\label{sssec:SFHmodel}

Some authors have argued for epochs of
enhanced star formation in the thin disk \citep{Rocha2000,Marcos2004}.
The errors in determining the age of
the disk, however, are larger than the peaks and troughs in the SFH. We therefore 
approximate these SFHs by a smooth function.  
The two most common forms of a smooth SFH of the
entire Galaxy are the exponential form \citep{MacArthur2004}:

\begin{equation}
\label{exp.eqn}
\Psi (t) = k \> exp\Biggl[\frac{-(t-t_{0})}{\tau}\Biggr]
\end{equation}

\noindent and the equation used by \citet{Sandage1986}:

\begin{equation}
\label{sandage.eqn}
\Psi (t) = k \> (t-t_{0}) \> exp\Biggl[\frac{-(t-t_{0})^2}{2\tau^2}\Biggr],
\end{equation}

\noindent where $t_{0}$ (only in our adaptation of the equation)
is the time when star formation started in a given population.
The factor $\tau$ determines the shape of the distribution
of stellar ages. 
It has been accepted that the bulge began to form shortly after
the initial spheroid collapse \citep{Terndrup1988, Ortolani1996, Idiart1996}.

Since the spheroid is the oldest galactic component, starting to form 13~Gyr ago at $t=0$
and the thick disk started forming at $t=1.0$~Gyr, 
we will adopt $t_0$=0.5~Gyr for the bulge. 
The thin disk did not begin to form until $\sim$10 Gyr ago
\citep{Liu2000, Sandage2003, Peloso2005}.
Thus we adopt
$t_0=3.0$~Gyr for the thin disk.  We also set $t_{end}$ to be the time where star formation ends.
Its value is 4.5~Gyr for the spheroid, 6~Gyr for the thick disc, and 13~Gyr for both the thin disc and bulge (meaning that star formation is ongoing in those two galactic components).

In Eqn.~(\ref{exp.eqn}), $\tau$ corresponds to the time $t-t_{0}$ when an
average-aged star
was formed, if $\tau$ is positive and star formation is allowed to continue
indefinitely into the future (i.e.,
star formation does not stop at $t_{end}$). In Eqn.~(\ref{sandage.eqn}),
$\tau$ is the time $t-t_{0}$ when the SFR peaks.
The values of $\tau$ for the 16 cases (two SFR representations for all components combined with the three average ages of the thin disc and bulge)
are listed in Table~\ref{tab:SFR}
The values of $t_0$, the assumed mean ages, and $t_{end}$ of the four galactic components are also listed in the table.

The two equations each better represent a different aspect of the SFR
in the Galaxy. Eq.~\ref{exp.eqn} is strictly a
decreasing or increasing SFR depending on the sign of $\tau$. It is rather unphysical 
since the SFR should not switch on instantly.
Eqn. (\ref{sandage.eqn})
correctly addresses the initial rapid increase of stellar formation and
then a decline after a certain peak.
However, the
decline of the SFR dictated by Eq.~\ref{sandage.eqn} is too rapid and does not account for the
small but non-negligible population of young and intermediate age stars
observed in the bulge.

\begin{figure}
\centering  
\includegraphics[angle=0, clip, width=3.6in]{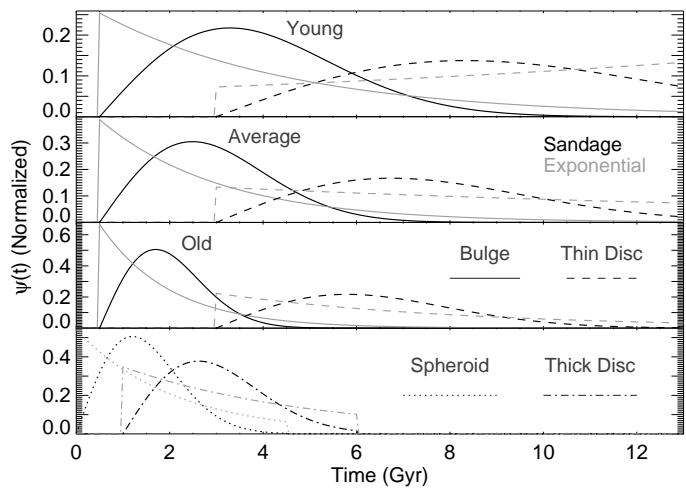}
\caption{Star formation history of the four components of the Galaxy assuming three different mean ages 
(called ``Young", ``Average" and ``Old" in Table~\ref{tab:SFR}) and
two different SFR functions for the thin disk and bulge
(the ``Exponential" and the ``Sandage" function, Eqs.~\ref{exp.eqn} and \ref{sandage.eqn}, respectively).}
\label{SFH.fig}
\end{figure}

Given the above considerations, 
we consider six SFH functions each for the bulge and thin disc and two SFH functions each
for the spheroid and thick disc (i.e., we  model the SFH
using the two equations and the three average mean age values
listed above). 
These models are plotted in Fig.~\ref{SFH.fig}  with all components normalized to unity.  Notice how the SFHs of the thin disk according to
Eqn. (\ref{exp.eqn}) can be either decreasing, uniform, or increasing due to the
uncertainty in determining the average age of the thin disk.  

\subsection{Metallicities}
\label{ssec:metallicities}

\begin{figure}
\centering  
\includegraphics[angle=90,width=3.6in]{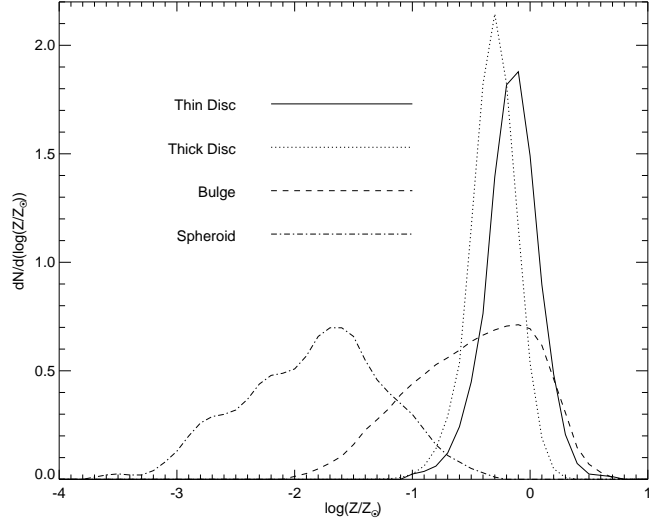}
\caption{Adopted metallicity distributions for the different parts of the Galaxy.}
\label{MetalDist.fig}
\end{figure}

Accounting for
the metallicity distribution of the Galaxy, $\beta(z)$, and how it has changed over time, $\zeta(t)$,
will reduce the uncertainty in the calculated stellar lifetimes (\S~\ref{ssec:lifetimes})
and PN visibility times (\S~\ref{ssec:PNLifetimes}).
In Fig.~\ref{MetalDist.fig}, we display the adopted metallicity distributions expressed as a function of $Z$:

\begin{equation}
\beta(Z) = {dN \over d(\log(Z/Z_{\odot}))}. 
\end{equation}

Metallicity distributions, expressed in terms of 
[Fe/H] $=\log($Fe/H$) - \log($Fe/H$)_\odot$, are converted to $Z$ using the following formulae
\citep[and references therein]{Nissen1991}:

\begin{equation}
\label{metalconvert.eqn}
\log(Z/Z_\odot) = 
\cases{0.60[$Fe/H$] &  $\;$ for [Fe/H]  $\ge$ -1.0 \cr
0.4 + [$Fe/H$] & $\;$ for [Fe/H] $<$ -1.0 \cr}.
\end{equation}

The stellar population of the spheroid has the lowest metallicities in the Galaxy
($-2.0 < $ [Fe/H] $< -1.5$; \citealp{Ryan1991, Marquez1994, Carney1996, Beers2005}).
In this work, we use the prescription of \citet[their solid line in Fig.~16]{Carney1996},
who already express their distribution function in terms of $Z$ instead of [Fe/H].
The metallicity of the bulge peaks at [Fe/H]$\sim$0.2~dex,
\citep{McWilliam1994, Minniti1995, Sadler1996, Rich1998,
Ramirez2000, Feltzing2000, Zoccali2003}.  We use the
continuous [Fe/H] distribution obtained by \citet[their Fig. 4]{Minniti1995} and apply
Eqn.~(\ref{metalconvert.eqn}) to convert to a $Z$ distribution.
The metallicity of the thick disk ranges between
$-0.6 < $[Fe/H]$ < -0.4$, with a dispersion of 0.2--0.4 dex
\citep{Nissen1991,Soubiran2003,Bartasiute2003, Schuster2005}.  We model
the [Fe/H] distribution of the thick disk as a Gaussian with a mean value of --0.5
and $\sigma = $ 0.3 dex.  
The thin disk has the highest metallicities peaking near solar metallicity with a small
dispersion of only $\sim$0.25 dex \citep{Bartasiute2003,Soubiran2003,Pont2004,
Nordstrom2004,Haywood2005}.  We use the distributions of
\citet[their Fig. 9]{Nordstrom2004}, who express metallicity in terms of $Z$.
Also, the metallicity distributions for the four galactic components, have been derived mainly from
G dwarfs, i.e., for main sequence stars; therefore, the distribution is not affected by the stellar evolution correction factor and
we do not have to apply $\delta$($\phi$) (Eq.~\ref{delta.eqn}).

\subsubsection{The Age-Metallicity Relation}
\label{sssec:AMR}

We have assumed that there is an AMR
in all four components of the Galaxy, represented by a geometric equation.  The average
metallicity of a star formed at time $t$ is:

\begin{equation}
\langle \zeta \rangle (t) = \log(Z/Z_{\odot}) = a {(t-b)}^c+d,
\end{equation}

\noindent with fitting parameters $a$, $b$, $c$ and $d$.  Several authors
\citep{Carraro1998,Ibukiyama2002,Bensby2004,Ibukiyama2004,Haywood2005}
however, have found scattering in the AMR, with $\sim90\%$
of stars formed at a given time scatter over about 1~dex interval in metallicity.  
We will therefore assume that 
the metallicity distribution of stars formed at any given time $t$ follows a Gaussian function:

\begin{equation}
\zeta(t,\log(Z/Z_{\odot})) = {1 \over \sqrt{2 \sigma}} \exp\Biggl[{-( \langle \zeta \rangle(t) -\log(Z/Z_{\odot}))^2 \over 2 \sigma^2}\Biggr],
\end{equation}

\noindent where $\sigma$, the width of the AMR at any given time $t$, is chosen to be 0.25 dex for the bulge and spheroid and 0.15 dex
for the thin and thick disks, since $\Delta \log(Z/Z_\odot) = 0.6 \Delta $[Fe/H] for these higher metallicity components.  

The fitting parameters are determined by minimizing the equation:

\begin{equation}
\chi^2 = \sum  (\beta_{i} - \sum \Psi(t_j) \times \zeta(t_j,\log(Z/Z_{\odot})_i))^2,
\end{equation}

\noindent where $i$ is in steps of 0.1~dex from $\log(Z/Z_{\odot})$=--4.0 to 2.0 dex, $j$ is in steps of 0.1~Gyr from 0 to 13 Gyr,
and the SFR, $\Psi(t)$, is normalized such that $\int_{0~{\rm Gyr}}^{13~{\rm Gyr}} \Psi(t) dt = 1$.

\begin{figure}
\centering  
\includegraphics[width=3.6in,angle=90]{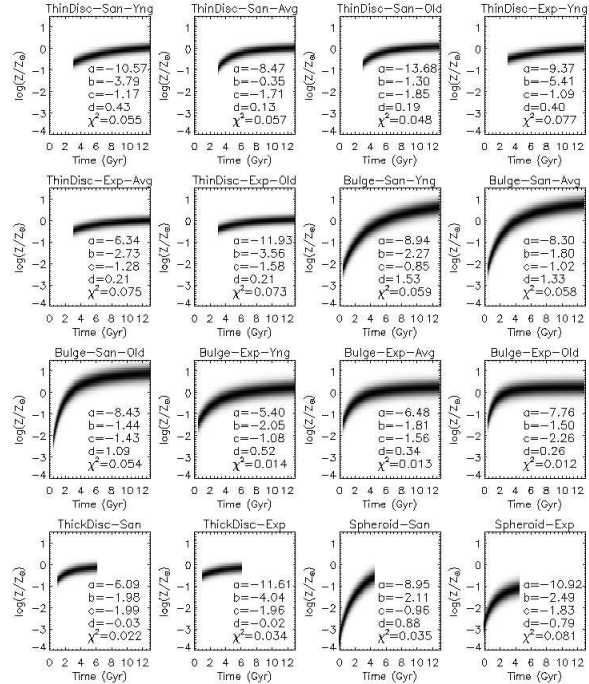}
\caption{The age-metallicity relation for the four galactic components.
For the thin disk and the bulge the relation is shown for the exponential
(marked ``Exp" in the panels' titles) and Sandage (marked ``San" in the panels'
titles) forms of the equation (Eqns.~\ref{exp.eqn} and \ref{sandage.eqn}, respectively),
as well as for the Old, Average and Young ages of the
populations (see Table~\ref{tab:SFR}).}
\label{AMR.fig}
\end{figure}

In Fig.~\ref{AMR.fig} we display the AMR. The 16 panels depict the AMR for the four galactic components,
where for the thin disk and bulge two different SFR equations (Eqs.~ \ref{exp.eqn} and \ref{sandage.eqn}) have been used 
each with three mean ages (see Table~\ref{tab:SFR}).
The greyscale indicates
the relative distributions of metallicities for a given epoch, but does not display the variability of the stellar formation
rate as a function of time.

Finally, we can write the equation to determine 
the scaling constant, $k$, used in the SFR Eqs.~(\ref{exp.eqn}-\ref{sandage.eqn}):

\begin{equation}
\label{k.eqn}
k = \frac{M}{\int_{0~{\rm Gyr}}^{13~{\rm Gyr}} \Psi(t) \delta(t, \xi, \tau_{*}) dt},
\end{equation}

\noindent where $M$ is the total current ``luminous" matter (\S~\ref{ssec:lifetimes}) of the galactic component 
under consideration (determined in Sec.~\ref{ssec:massMW}) and
$\tau_{*}$ is the stellar lifetime weighted by the AMR.  
  
\subsection{Stars that do not ascend the AGB due to early binary interactions}
\label{ssec:binaries}

For the present calculation, we assume that single stars and primary stars in binaries fitting the
criteria outlined in \S~\ref{ssec:IMF} to \ref{ssec:metallicities}, are candidates to make PNe. 
We implicitly assume that all of these systems ascend the AGB and that only post-AGB stars can make a PN.
The latter assumption is justified on the grounds that almost all of the PN central stars analyzed spectroscopically
are shown to be post-AGB stars \citep{Mendez1989,Napiwotzki1999}. 

Because of these considerations, we want to exclude from our PN progenitor sample those stars that do {\it not}
ascend the AGB. There are three reasons why a star might not go through the AGB phase. (i) If its helium core mass
remains smaller than 0.47~\msun\ the star will not ignite core helium and will instead become a He-WD right after the RGB phase 
\citep{Jimenez1996}. (ii) 
Stars that do ignite helium in the core
and develop a CO core on the horizontal branch, might never ascend the AGB if their post-RGB envelope mass is too small \citep{Dorman1993}. It is difficult to estimate exactly how many stars meet this criterion,
but \citet{Heber1986} estimates that the early horizontal branch channel (the stars with low envelope
masses) contribute only 2\% to the total number of WDs.  
(iii) Finally, post-RGB stars might be prevented from ascending the AGB if they suffered a common envelope
with a companion on the RGB, becoming horizontal branch short period binaries. 
These binaries consist of very blue  horizontal branch stars
(called subdwarf-B stars \citep{MoralesRueda2003}) with low-mass main sequence companions. Their very blue colors are due to particularly small envelope masses.
If so, just as is the case for stars in point (ii) above,  the primaries will not ascend the AGB. 

The first group is already excluded from our counting since we impose a lower mass limit below which post-AGB stars do not make
PNe (because their transition time between AGB and PN phase is too long - see \S~\ref{ssec:PNLifetimes}). In addition, the 0.47~\msun\ helium core mass cutoff, corresponds to 
a progenitor mass of $\sim$0.75~\msun\ \citep{Jimenez1996}, which is similar to the turnoff mass of the oldest galactic components;  this is like saying that the Galaxy is on the whole too young
for a sizable fraction of its stars to follow this evolutionary channel.  

The number of stars in the second group is very hard to estimate. It could be obtained
theoretically by integrating the mass-loss rates from the start to the end of the RGB to derive the actual distribution of
envelope masses for post-RGB stars and then use the \citet{Dorman1993} models to determine how many of these stars
do not ascend the AGB. We have however decided against taking this route because of the many arbitrary choices to be
made along the path, because only a few of the \citet{Dorman1993} tracks completely miss the AGB, and also
because \citet{Heber1986} argue that the size of the group is small.
 
We have however, taken into account the members of the third group.
We estimated the number of galactic systems that undergo a common envelope interaction on the RGB to be 
$\sim$10\% of all systems (but see Paper II for details). This number was determined from considering that 57\% of systems have
a stellar companion, 
25\% of which have separations $a<$600~\rsun, 68\% of which have mass ratios larger than
0.3 \citep{Duquennoy1991}. The separation and mass ratio requirements are due to the RGB radial expansion and tidal capture 
mechanisms \citep{Soker1996} and the minimum mass companion that will emerge from a common envelope on the RGB \citep{Terman1996,Sandquist1998,Yorke1995,DeMarco2003}.

Although these estimates are rather approximate, we feel reassured that the number of stars that
do have time to evolve off the main sequence but {\it do not} ascend the AGB is small. In our calculation we
reduce the number of PN progenitors by 10\% to account for this eventuality. More on this topic will be included in Paper II.

\subsection{The PN visibility time}
\label{ssec:PNLifetimes}

The number of PNe in a population depends directly on the time during which a PN is visible.
In first approximation, we could use a mean PN visibility time for all PNe.
Using a mean value is, however, problematic because
the PN visibility is not only a function of its kinematic properties, but also of 
the temperature and luminosity of its central star. We therefore calculate the PN visibility time taking into
account the time central stars of different masses take to heat up to ionization temperatures, as well as the time they take to fade enough that the PN cannot remain ionized. 
To this we add an additional criterion that no PN can remain detectable for more than 35\,000~yr
after its central star leaves the AGB, since its gas has by then become too dispersed. 

This upper limit of 35\,000 years is possibly the {\it single most complicated issue in our counting}. Changing this number directly alters the predicted galactic PN population. We stress however that the choice of 35\,000~yr, while arbitrary in some respect, was made to be able to compare the predicted galactic PN population with specific observational estimates (see below). {\it As a result, while both our prediction and the observationally-based estimates might be subject to large uncertainties, the comparison between the two has a much smaller uncertainty, and in fact could be considered independent of the PN visibility time that is the source of so much debate.}

We will only compare our predictions to observational estimates of the galactic PN population that are limited by a maximum radius of detectability of $\sim$0.9~pc (see \S~\ref{ssec:PNnumbers}). For PNe with radii larger than this threshold the surface brightness is so low that the count becomes incomplete.  The PN visibility time corresponding to a radius of 0.9~pc is 35\,000~yr, because the average
expansion velocity of PNe is 25 \kms \citep{Phillips1989}. Hence, our choice of this value as the maximum kinetic age of a PN.  

Central star luminosities have been determined 
by studying Large and Small Magellanic Cloud (LMC and SMC, respectively) PNe.
Values as low as $\log(L/L_{\odot}) = 0.9$ are derived, although most of them are brighter
than $\log(L/L_{\odot}) = 1.5$ \citep{Dopita1997, Jacoby1993}. We therefore adopt
$\log(L/L_{\odot}) = 1.5$, as the central star luminosity limit, below which the 
PN cannot remain ionized. For example, a central star descending from a 5.0-\msun\ progenitor, with quarter-solar metallicity (Z=0.004), 
heats and fades below $\log(L/L_{\odot}) = 1.5$ in $\sim$25\,000~yr \citep{Vassiliadis1994}. 
Its PN visibility time would therefore be limited to $\sim$25\,000~yr: when the
star fades, the PN, still with relatively high electron densities, will recombine quickly and
disappear. 
It was noted by the referee that the lower luminosity limit of LMC PNe could be due to  a limiting magnitude problem in the relatively far away galaxy. We note, however, that because of the relatively low maximum PN visibility time value adopted above (35\,0000 yr), very few central stars in our sample actually fade below $\log(L/L_{\odot}) = 1.5$. As a result, this limit has almost no effect on the predicted total galactic PN number.

Our second criterion is that no stars with an effective temperature lower than 25\,000~K 
($\log T_{eff} < 4.4$) will ionize their PNe.
For example, a central star from an 0.8-\msun\ progenitor takes a long $10^5$~yr to heat
to $\log T_{eff} = 4.4$ \citep{Schoenberner1983} and
by the time the star is hot enough, the circumstellar gas that would become the visible PN is likely to have 
dispersed enough that the PN will never be detected. From this criterion it follows that there is
a cut-off central star mass, below which a post-AGB star cannot produce a PN.
This cut-off value adds a source of uncertainty to the total number of PNe. The only evolutionary calculation 
for main sequence masses smaller than 1.0~\msun\ are those of \citet{Schoenberner1983}. From these we determined 
a cutoff central star mass
for PN production of 0.90~\msun. However, since the exact choice of 
this number influences the total number of PNe derived (but see \S~\ref{ssec:uncertainties}),
we also carried out the calculation with two additional values:  
0.85 and 0.95~\msun.  
In conclusion, the PN visibility time is the time taken by a central star to evolve between $\log T_{eff} = 4.4$ and
$\log(L/L_{\odot}) = 1.5$, with an upper limit of 35\,000~yr, minus the time the central star takes to reach $\log T_{eff} = 4.4$.

Post-AGB evolutionary times are also metallicity dependent.
\citet{Vassiliadis1994} show that
a 5.0-$M_{\odot}$ solar metallicity progenitor takes 70\,000 years to fade below $log(L/L_{\odot}) = 1.5$,
nearly three times longer than a central star from a quarter solar progenitor of the same mass. The PN of the former star will therefore
be visible for longer than the PN of the latter.
We therefore will run two sets of models. In the first, we use the evolutionary timescales of the 
metallicity-independent H-burning central star tracks of \citet{Bloecker1995} and \citet{Schoenberner1983},
while in the second, we use the metallicity-dependent timescales of the H-burning models of \citet{Vassiliadis1994}.  

\begin{figure}
\centering
\includegraphics[width=3.6in,angle=90]{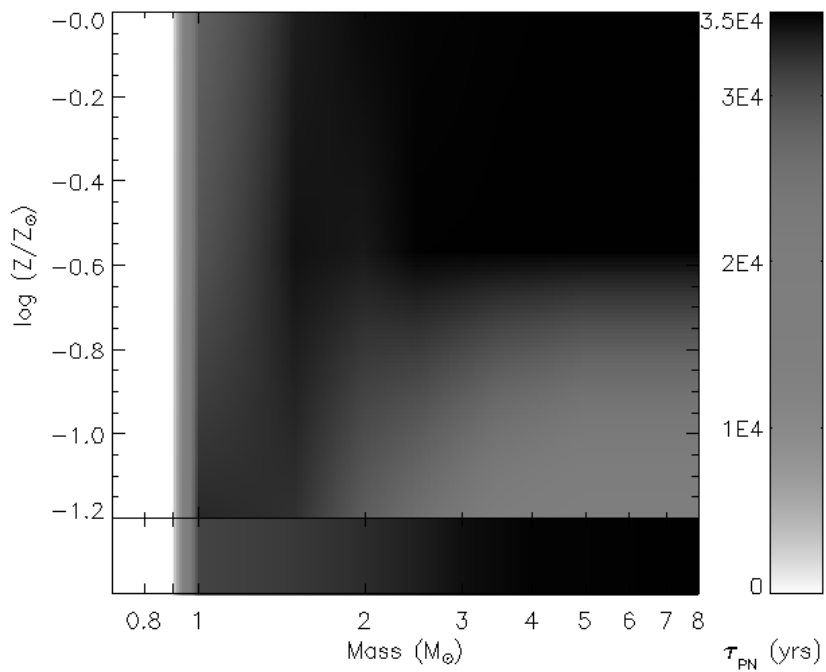}
\caption{Greyscale of the PN visibility time as a function of mass and metallicity
(top panel from the calculation of \citet{Vassiliadis1994}) and mass only
(bottom panel, from the calculation of \citet{Bloecker1995}.}
\label{PNLifetime15.fig}
\end{figure}

In Fig.~\ref{PNLifetime15.fig} we show a greyscale of the PN visibility time as a function of main sequence star mass
and metallicity for the \citet{Vassiliadis1994} tracks and as a function of mass only for the \citet{Bloecker1995} and 
\citet{Schoenberner1983} tracks. These have been calculated using a linear interpolation with respect to mass and metallicity, 
respectively. PN lifetimes of progenitors with masses not included within 
the range of the evolutionary models, are given PN visibility times corresponding to the boundary
values.  It is clear from this figure that the \citet{Bloecker1995} tracks predict longer
visibility times for higher mass
progenitors than do the \citet{Vassiliadis1994} models. This is mostly due to the different initial-to-final mass relation 
derived by the two models, where the \citet{Vassiliadis1994} tracks predict larger PN central stars masses than the
\citet{Bloecker1995} curves, in particular for higher mass progenitors. These more massive nuclei evolve faster and fade faster, so that
the PN visibility times are relatively shorter. Fig.~4 of \citet{Bloecker1995} shows that
their tracks predict an initial-to-final mass relation which is closer to the observationally-based relation
of \citet{Weidemann1987}. However Fig.~3 of \citet{Weidemann2000} shows that the updated observational
initial-to-final mass relation is actually closer to the determination of \citet{Vassiliadis1994}, except
for progenitor masses smaller than 2~\msun, where the observational relation is in between the \citet{Bloecker1995} and
\citet{Vassiliadis1994} predictions. Since more of the PNe visible today derive from progenitors with masses
smaller than $\sim$2~\msun, it is appropriate to adopt both sets of tracks and average the results.

\section{Results: the number and birth-rate of PNe and WDs in the Galaxy}
\label{sec:results}

To determine the total number of galactic PNe, we have considered
three possible IMFs (\S~\ref{ssec:IMF} and Fig.~\ref{IMF.fig}), three values for the galactic mass (\S~\ref{ssec:massMW};
Table~\ref{tab:SFR}), 
two SFH functions, three
mean galactic ages (\S~\ref{ssec:SFH} and Fig.~\ref{SFH.fig}), three central star mass lower limits
to make a PN and two PN visibility time models 
(\S~\ref{ssec:PNLifetimes} and Fig.~\ref{PNLifetime15.fig}).  
We therefore computed a total of 324 models and obtained 324 values for the total number of galactic PNe.
The range in these values corresponds to the cumulative uncertainty contributed by these assumptions.  

\begin{figure}
\centering
\includegraphics[width=3.6in,angle=90]{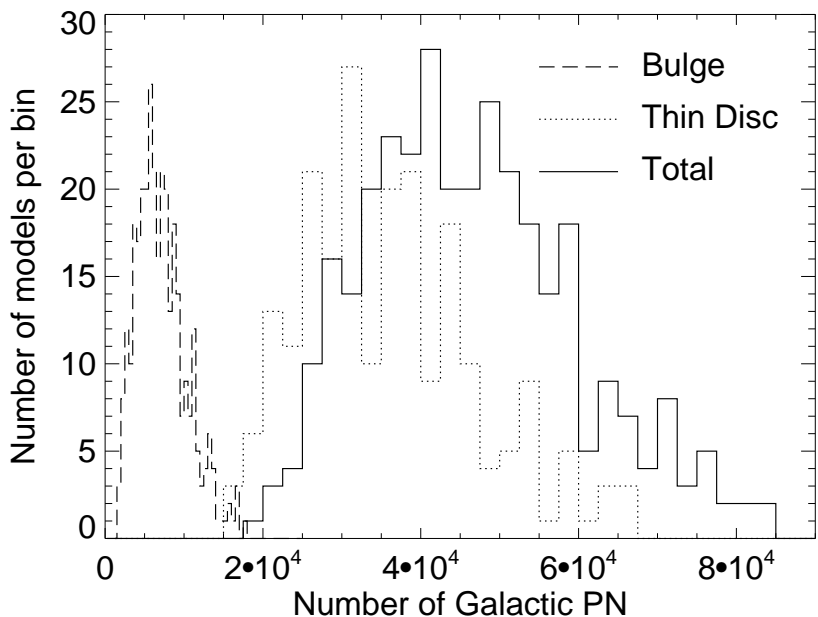}
\caption{Histogram of the PN population sizes predicted by different models.  
The bin sizes are 2500, 500 and 2500 for the bulge, thin disk and total respectively.}
\label{totpn.fig}
\end{figure}

The mean total number of PNe in the Galaxy with radii less than 0.9~pc 
is $(4.6\pm1.3)\times 10^4$ objects, with the
thin disk, bulge, thick disk and spheroid each contributing $(3.6\pm1.1)\times 10^4$, $7200\pm3200$, $1100\pm500$
and $70\pm80$ PNe, respectively.  The average visibility time for PNe with radii less than 0.9~pc is $26\,000\pm 3\,000$~yr.
The errors given represent the 1 $\sigma$ deviation in the values predicted by the models.  
Histograms of the predicted PN populations
for the thin disk, bulge and entire Galaxy are plotted in Fig.~\ref{totpn.fig},
where we show, visually, the uncertainty in the prediction due to the range of possible
assumptions (further discussed in \S~\ref{ssec:uncertainties}). Only one of our 324 models produces fewer than 20\,000~PNe, this implies that there should be more than 20\,000~PNe in the Galaxy from single stars and binaries at the 3$\sigma$ level.
It is interesting to notice that although the thin
disk is only $\sim$1.8 times more massive than the bulge, it produces $\sim$5 times more PNe. 

The galactic PN population derived by our model is lower than that derived
from the back of the envelope calculation (\citealt{DeMarco2005}, 
\S~\ref{sec:BotE}).  This is mainly due to the fact that simple averages for the age of the Galaxy as well as taking lifetimes and masses
of a  prototypical progenitor star, instead of using bins, biased the calculation toward larger PN populations.  

\begin{figure}
\centering
\includegraphics[width=3.6in,angle=90]{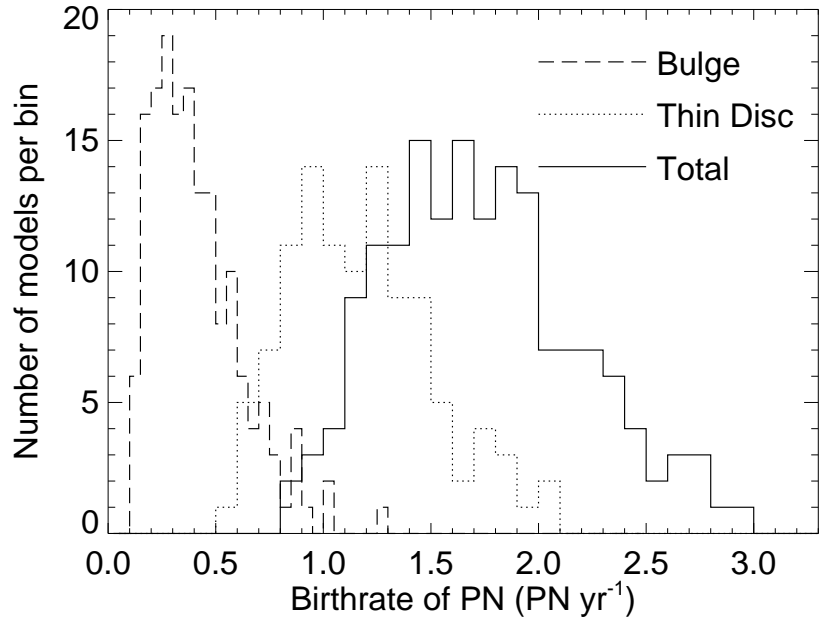}
\caption{Histogram of the PN
birthrates predicted by different models.   
The bin sizes are 0.05, 0.1 and 0.15~PN~yr$^{-1}$ for the bulge, thin disk and total respectively.}
\label{birthrate.fig}
\end{figure}

Quoting the PN birthrate
avoids the uncertainty related to the PN visibility time.
We find the birthrate of PNe in the 
Milky Way to be ($1.7\pm0.3$)~ PN~yr$^{-1}$, with the thin disk, bulge, thick disk and spheroid each contributing ($1.2\pm0.3$),
($0.4\pm0.2$), ($0.07\pm0.03$) and ($0.01\pm0.01$)~PN~yr$^{-1}$, respectively.  
Histograms of the predicted PN birthrates for the thin disk, bulge and the entire Galaxy are plotted in
Fig.~\ref{birthrate.fig}.

\begin{figure}
\centering
\includegraphics[width=3.6in,angle=90]{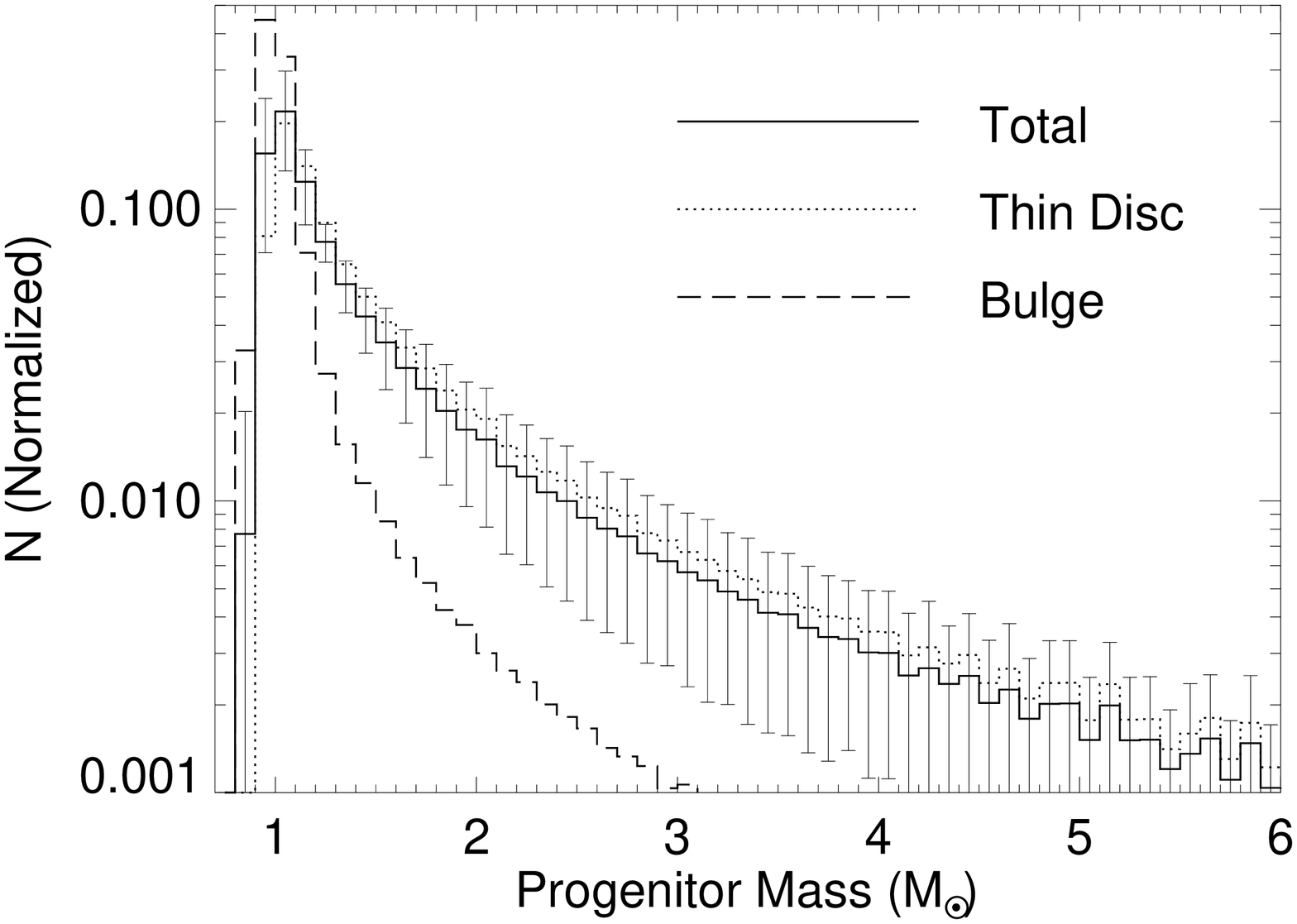}
\caption{PN central star progenitor mass distribution. Error bars are for the total population.}
\label{pnmassdist.fig}
\end{figure}

\begin{figure}
\centering
\includegraphics[width=3.6in,angle=90]{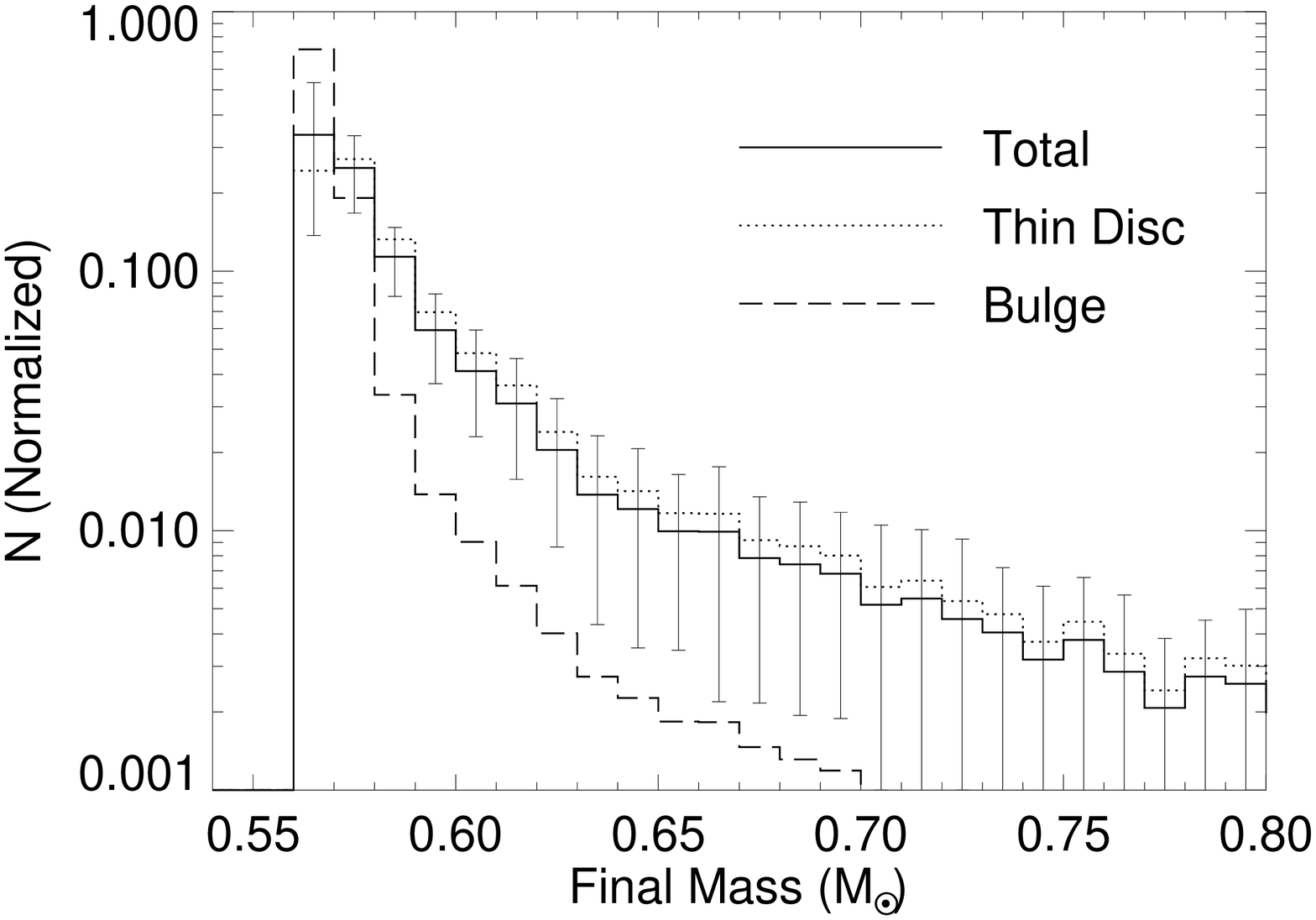}
\caption{PN central star mass distribution. Error bars are for the total population.}
\label{pnfinalmassdist.fig}
\end{figure}

The central star progenitor mass distributions is plotted in Fig.~\ref{pnmassdist.fig} for thin disk, bulge and the entire Galaxy, along with the 1 $\sigma$ errors for each bin. The mean and median progenitor masses are ($1.7\pm0.3$) and ($1.2\pm0.2$)~\msun, respectively. We can also use this distribution to check for consistency between the frequency of Type I PNe (21\%; \citet{Kingsburgh1994}) and the frequency of their progenitors. Type I PNe are those where the N/O$>$0.8 because of hot bottom conversion of carbon to nitrogen. This is only thought to occur in stars with main sequence mass larger than 4~\msun\ \citep{Becker1980}. We predict that ($6\pm4$)\% of the PN progenitors' masses are above 4.0~\msun, three times less than the observed Type I a frequency. The primary contribution to this error is the SFH; younger Galaxy models give higher mass progenitors while older Galaxy models give lower mass progenitors (see \S~\ref{ssec:uncertainties}). Using the initial-to-final mass relation from the \citet{Schoenberner1983}, \citet{Bloecker1995} and \citet{Vassiliadis1994} models we derived the central star mass distribution (Fig.~\ref{pnfinalmassdist.fig}). The mean and median central star masses are ($0.61\pm0.02$) and ($0.57\pm0.01$)~\msun, respectively.
%with only ($5\pm3$)\% of the values residing above 0.80~\msun.  

\begin{figure}
\centering
\includegraphics[width=3.6in,angle=90]{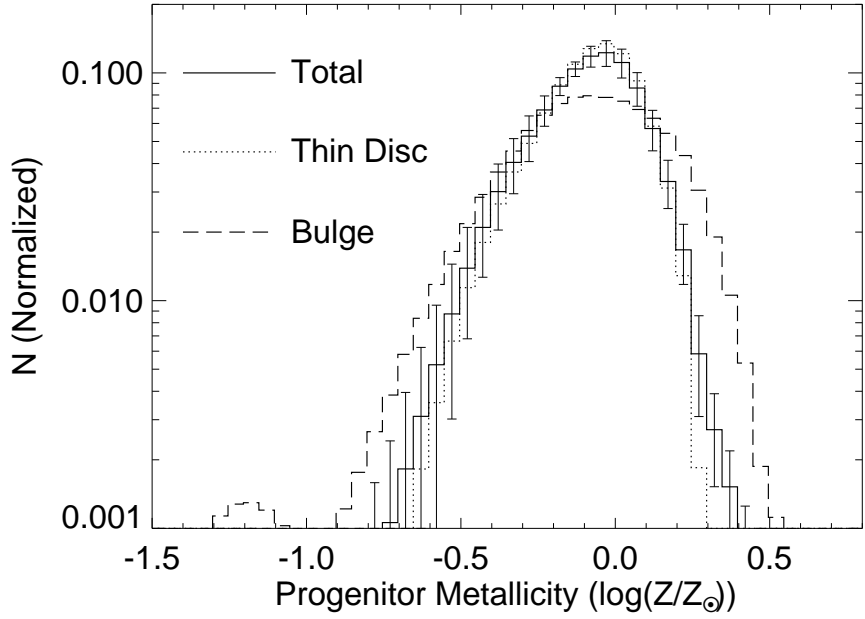}
\caption{PN central star 
progenitor metallicity distribution. Error bars are for the total population.}
\label{pnzdist.fig}
\end{figure}

The metallicity distribution of PN projenitors is shown in Fig.~\ref{pnzdist.fig}. Despite the large range of metallicities of stars formed in our Galaxy, the PNe we see today are mainly confined to progenitors that had near solar metallicity. The metallicity distribution for the entire Galaxy can be fitted by a Gaussian curve with a mean of --0.07~dex and a dispersion of 0.16~dex. Again, by looking at the AMRs, all components had reached an average metallicity greater than $log(Z/Z_{\odot})=$--1.0 by $t \sim 1.5$~Gyr, when the lowest mass PN progenitors formed.  

\begin{figure}
\centering
\includegraphics[width=3.6in,angle=90]{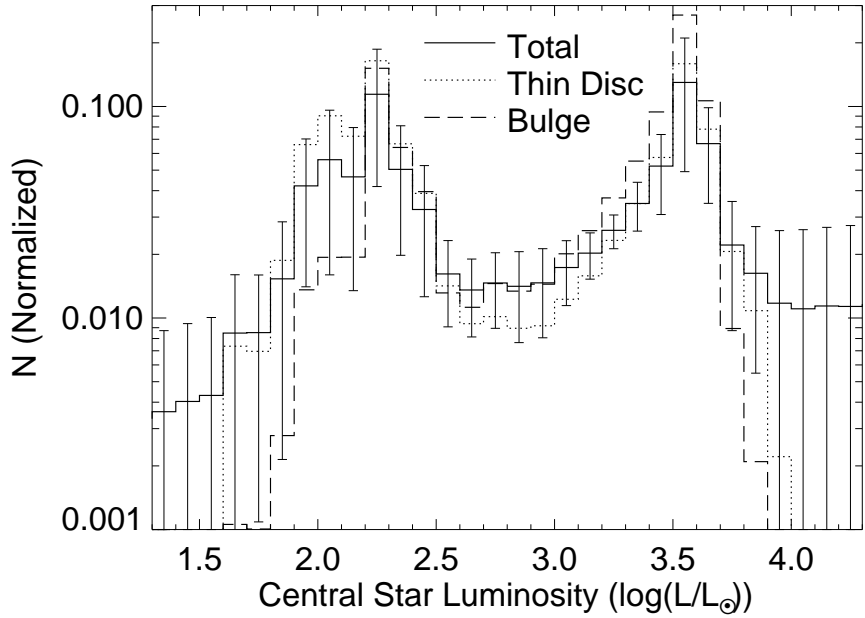}
\caption{Central star luminosity function of PN.}
\label{pnlumdist.fig}
\end{figure}

Finally, by determining the PN visibility time associated with each central star mass, 
we computed (Fig.~\ref{pnlumdist.fig}) the expected luminosity distribution of our galactic central stars.
The median central star luminosity is $\log(L/L_{\odot})=2.3\pm0.2$. The bimodal nature of the luminosity distribution (with peaks at $\log(L/L_{\odot})=2.2$ and 3.5), is due to the fact that lower mass central stars (which are the most numerous) have a slow luminosity evolution right after the AGB (the 0.569-\msun\ solar metallicity central star of \citet{Vassiliadis1994} spends 27\,000~yr at approximately $\log(L/L_{\odot})=3.5$), followed by a speeding up of the evolutionary time (it takes only 13\,000~yr for its luminosity to drop to $\log(L/L_{\odot})=2.4$), later followed by a slowing down of the evolutionary time (it takes another 60\,000~yr for the stellar luminosity to drop further to $\log(L/L_{\odot})=2.1$). We wonder here whether the unexplained bimodal behavior of the PN luminosity function (PNLF) of external galaxies (e.g., see the SMC PNLF of \citealt{Jacoby2002}) is due in part to this effect.

The predicted  post-AGB WD birthrate for the Galaxy is ($2.4\pm0.5$)~WD~yr$^{-1}$ with the thin disk, bulge,
thick disk and spheroid each contributing ($1.3\pm0.4$), ($0.7\pm0.3$), ($0.09\pm0.02$) and ($0.14\pm0.04$)~WD~yr$^{-1}$, respectively.  Thus, the ratio of the PN birthrate to the post-AGB WD birthrate for the Galaxy is  ($0.73\pm0.10$), while, for the thin disk, bulge, thick disk and spheroid  it is ($0.89\pm0.01$), ($0.58\pm0.21$), ($0.71\pm0.22$), and ($0.08\pm0.08$), respectively. The small ratios in the older galactic components are due to the particularly long times taken by low mass progenitors to reach $\log (T_{eff}/$K$) = 4.4$ and the consequent low PN production rate.

\subsection{The uncertainties}
\label{ssec:uncertainties}

Here we analyze the main sources of uncertainty incurred because of our six assumptions. The 324 models produce 324 values of the PN population. These models can be divided into groups within which a given assumption is kept constant. For instance, there will be two groups, each containing 162 models, where one groups is calculated using the exponential equation (Eq.~\ref{exp.eqn}) for the SFR and the other using the Sandage equation (Eq.~\ref{sandage.eqn}). One can then compare pairs of PN population predictions obtained by keeping all parameters constant {\it except} for the SFR equation. From these 162 pairs one can extract 162 percentage differences. The mean and standard deviation of the 162 percentage differences is reported in Table~\ref{tab:Error}.  

The main source of error is from the uncertainty on the age of the Galaxy, primarily that of the thin disk. The younger the thin disk, the higher the SFR at epochs when the turn-off mass was above 1.0~\msun, the longer the PN lifetimes, the more PNe are predicted. The second main contribution to the error is the uncertainty in the mass of the Galaxy.  Our adopted value for the Galactic mass of ($6.2\pm1.1)\times 10^{10}$~\msun\ (\S~\ref{ssec:massMW}), is much lower than the value of 100 billion~\msun\ often found in the literature \citep[e.g.,][]{Innanen1966}.  We believe that this higher mass figure should be the {\it number} of stars in the Galaxy, not the {\it total mass} of stars in the Galaxy.  Indeed, we calculate that there are $1.3\pm0.2\times 10^{11}$ stars in the Galaxy, from which an average galactic star has a mass of $\sim$0.5 \msun\ (as can be calculated by taking the mean of any IMF). The IMF contributes the next largest error, where the \citet{Kroupa2001} disk IMF contributes the least amount of PNe, while the \citet{Chabrier2003small,Chabrier2003big} disk IMF produces the most.   Finally, the central star mass cut-off value below which no PN is formed, contributes the fourth largest source of uncertainty. The reason for this relatively small contribution to the uncertainty by a parameter which, after all,
plays a large role at the low mass end of the IMF (where many more stars reside), is that it only plays a prominent role in old populations, which are not the main contributors to the total number of PNe in the Galaxy.

We also point out the importance of using metallicity-dependent stellar lifetimes and of using the AMR. If we had used exclusively solar-metallicity stellar lifetimes, the turn-off mass of the Galaxy would have been 0.94~\msun\ and the ratio of the PN formation rate to the WD formation rate would be $\sim$0.90 (instead of 0.73) for the entire Galaxy.  However, since metallicities were actually low at the beginning of the formation of the Galaxy, and since lower metallicity stars of 1~\msun\ live shorter than the same stars at higher metallicities, then the turn-off mass of the Galaxy is reduced to 0.74~\msun, which significantly reduces the ratio of the PN to WD formation rates for the older components.

\subsection{Comparison with Observations}

In this section we compare our prediction of the galactic PN population size, birthrates,
birthrate densities and other parameters with observations or observationally-based
estimates.

\subsubsection{PN numbers}
\label{ssec:PNnumbers}

Having predicted the number of galactic PNe with radii smaller than 0.9~pc, we need to compare it with the observed numbers.
The number of galactic PNe observed so far is $\sim$2000-2500 \citep{Parker2003}, but the {\it total} number is no doubt larger, since
distant PNe close to the galactic plane are systematically obscured by dust. 
Extrapolating the total number of PNe from the observed one, based exclusively on visibility arguments is extremely difficult. Estimates of the total galactic PN population based on local PN density are tied to the controversial distance scale.
An alternative estimate of the number of galactic PNe has been obtained by \citet{Jacoby1980} and \citet{Peimbert1990,Peimbert1993},
by using the PN luminosity function (PNLF) for several extragalactic PN populations. We adopt these two estimates for our comparison. {\it We do not claim that these observationally-based estimates are better than many others found in the literature, but we do claim that these estimates have very clearly defined biases. By applying the same biases to our theoretical prediction we can achieve a much more meaningful comparison, even in view of the many uncertainties.} 

The method of \citet{Peimbert1990,Peimbert1993}  counts the brightest PNe in a given galaxy (estimated to be within 0.8-2.5~mag of the brightest).
It then assumes that the PNLF for
that galaxy is the same as the PNLF determined by \citet{Jacoby1989a}, \citet{Jacoby1989b} and \citet{Ciardullo1989} and extrapolates
the observed number of bright PNe to
8~mag down the PNLF. This figure was adopted because it was thought to include the faintest PNe \citep{Jacoby1980}. In this way, 
the total number of PNe in {\it that} galaxy is estimated. By assuming a PN visibility time of 25\,000~yr,
the PN birthrate can then be derived (PN~yr$^{-1}$),
and using that galaxy's bolometric luminosity, the PN birth rate per unit luminosity is calculated (called $\dot \xi$). 
The PN birth rate per unit luminosity, $\dot \xi$, is derived in this way for several galaxies and then plotted 
against each galaxy's bolometric luminosity as well as its intrinsic $(B-V)_0$ color index.
These relationships are then used, together with our own Galaxy's
bolometric
luminosity and color index, to derive $\dot \xi$ for our Galaxy. This number is then 
converted back to an absolute number of PNe using our Galaxy's bolometric luminosity and the
same PN visibility time. 
%Besides, the PN visibility time of 25,000 yr adopted by Peimbert is similar to the visibility time of 26,000 yr for PNe with radii less than 0.9 pc.  
Using this technique, \citet{Peimbert1990,Peimbert1993}
estimated the total number of galactic PNe to be $7\,200\pm1\,800$.  \citet{Jacoby1980} used a very similar technique to derive a total number of $10\,000\pm4\,000$ PNe with magnitudes brighter than 8~mag below the PNLF's bright end cutoff.  

The total galactic PN population derived in this way is technically independent of the
PN visibility time adopted (since its value is first used in a multiplicative way, and then divided out). However, a dependence on the visibility time returns when extrapolating the total PN populations to 8~mag down the PNLF. Today's deep surveys (e.g., Parker et al. 2003) have made it clear that 8 mag is not enough for a complete census. The reason why Jacoby (1980) chose this limit was that the \citet{Abell1966} PNe, then thought to be some of the largest and faintest, have radii as large as $\sim0.9\pm0.1$~pc depending on the distance scale used \citep{Steene1995,Zhang1995,Phillips2004,Phillips2005}.  Using the fact that the mean expansion rate of a PNe is 25 \kms, a maximum radius of detectability of 0.9 pc corresponds to a maximum visibility time of 35\,000 years.  This kinematic age is smaller than that of the largest PNe known today. However, by adopting the same limiting age, we have insured that our theoretical prediction is compatible with the total galactic PN population estimated by Peimbert (1990,1993) and Jacoby(1980).

%Recently, Frew (priv. comm.) used the most up to date distance scale to show that there are 28\,000 PNe in the Galaxy with radii less than 1.5 pc based on observations of nearby PNe  Assuming that the radius of a PNe is directly proportional to its age, then there should be 0.9/1.5 $\times$ 28\,000 = 16\,800 PNe with radii less than 0.9 pc.  He quoted an error of 15\% in the distance scale, implying the total error in the number of PNe is 32\% since his method used the column densities of PNe in the galactic disc which in turn depends on the square of the distance scale.  Thus one would expect $16\,800 \pm 5\,400$ PNe with radii less than 0.9 pc using this method.  

Calculating the weighted average assuming Gaussian statistics of the two independent measurements of $7\,200\pm1\,800$ and $10\,000\pm4\,000$,  yields $8\,000\pm2\,000$.  
 {\it This estimate is about six times smaller than our estimate and well outside the derived error bars at the 2.9$\sigma$ level.} Due to the unprecedented care we put in estimating {\it all} sources of uncertainty, and to the fact that when in doubt we have likely erred on the side of prudence, we confidently suggest that the difference between the predicted and observationally-estimated numbers is significant. {\it In line with the reasoning above, it is therefore meaningless to compare our theoretical estimate with total PN numbers determined with other methods.}

\subsubsection{PN birthrates}
\label{ssec:PNbirthrates}

We can also compare our predicted  PN birthrate density with the PN birthrate densities predictions from local counts. Both predictions and observational determinations have their own distinct caveats. Our theoretical prediction of the PN birthrate density is independent of the adopted PN visibility time, but is dependent on the adopted model of galactic density (see below). The observational value depends on the as yet uncertain PN distance scale (to the fourth power!), the luminosity model of the Galaxy as well as the PN visibility time which is used to determine the birthrate from the local PN density.

To compute birthrate {\it densities} (PN~yr$^{-1}$~pc$^{-3}$) from our PN birthrates (PN~yr$^{-1}$) we adopt the density equations for the thin disk by \citet{Drimmel2001,Robin2003} and \citet{Naab2006} and average the density over a region between 7.5
and 8.5 kpc from the galactic center and 0.32~kpc above and below the galactic plane, the average mass in a local cubic parsec
is estimated to be $(9.0 \pm 2.0) \times 10^{-13}$ times that of the entire thin disk (where the error
comes from the standard deviation on the mean of the three density equations used).  
Thus the local PN birthrate density (using the thin disk PN birthrate value) is (($1.2\pm0.3$)~PN~yr$^{-1}) \times 
$(($9.0 \pm 2.0) \times 10^{-13}$~pc$^{-3}$) = ($1.08\pm0.46)\times10^{-12}$~yr$^{-1}$~pc$^{-3}$, which is lower, but within our quoted error, than the 
value of $3 \times 10^{-12}$~yr$^{-1}$~pc$^{-3}$ by \citet{Pottasch1996} or the value of 
$2.1 \times 10^{-12}$~yr$^{-1}$~pc$^{-3}$ by \citet{Phillips2002} (see also other estimates in the summary by  \citealt{Phillips1989} and \citealt{Phillips2002}).

%The total PN galactic population determined using local PN density cited above use local and galactic luminosity arguments to extrapolate the local population to the entire Galaxy. These arguments are very different from the density argument we use to convert our predicted galactic population to a local PN density. This, is one of the reasons why  PN birthrate densities estimated from local counts are {\it larger} than our prediction, while total galactic PN populations estimated from the {\it same} local counts are {\it smaller} than our prediction. Other reasons for the discrepancy are that the PN birthrate densities and total galactic PN populations estimated from local counts rely on the observed local PN density which scales as the fourth power of the adopted distance scale.

Finally, we should remark that the predicted galactic PN population size (and PN birthrate density) that derive from 
binary interactions only (Paper~II), will obviously be smaller than we predict assuming that also single stars make PNe.
While the total PN population size will then be more in line with what is estimated from observations, the smaller
PN birthrate densities will be at odds with the estimates
based on local PN counts. The only way to reconcile our predictions of the birthrate densities with 
observationally-based counts will then be to admit that the
galactic distance scale should be revised to overall larger distances (which will reduce the local observed PN density).
This discussion will be expanded in Paper~II.

\subsubsection{PN central star masses}

We can also compare our theoretically-predicted mean central star mass and mass distribution to those from observed samples. 

Galactic PN central star masses are derived in two ways. The first (photometric) method uses central star
apparent magnitudes, reddenings, effective temperatures and
distances to locate the central star on the Hertzsprung-Russell diagram. Masses are then derived by 
comparing these positions to those of theoretical evolutionary tracks.
The second (spectroscopic) method uses stellar atmosphere models to fit the central star spectra and derive effective temperatures
and gravities. The masses are derived by comparing these two parameters 
(which are distance independent) with theoretical evolutionary tracks on the $\log g$--$T_{\rm eff}$ plane.

From a sample of 76 galactic central stars, \citet{Gorny1997} derive central star masses using the photometric method.
Their mean mass is 0.624~\msun. The combined samples (24 central stars) 
of \citet{Rauch1998}, \citet{Rauch1999} and \citet{Napiwotzki1999},
who use the spectroscopic method, yield a mean mass of 0.594~\msun. 
Only three objects are common to the two samples and the mass
estimates are quite discrepant (NGC~6720: $M = 0.611$ vs. 0.56~\msun; NGC~7094: $M=0.583$ vs. 0.87~\msun; 
NGC~3587: $M = 0.854$ vs. 0.55~\msun for the former and latter methods, respectively), revealing the uncertainties
inherent to this mass determinations. 
Our predicted mean central star mass, (0.61$\pm$0.02)~\msun, falls in between the values obtained via the
photometric and spectroscopic methods. 
The minimum central stars mass in the
observed galactic samples (0.56~\msun) is the same as our prediction, supporting the
argument that there is a minimum central star mass to make a visible PN. The mass distribution of
the photometrically-derived sample is broader than ours (Fig.~\ref{CSmassComp.fig};
we do not pay much attention to the apparent bi-modality of the
histogram since that could be due to sampling), while that of the spectroscopically-derived sample is more similar to ours.

PNe in the LMC and SMC do not suffer from the distance uncertainties present for
galactic PNe and the photometric method can be applied with fewer caveats. 
\citet{Barlow1989} adopts a mean central stars mass for both galaxies of (0.586$\pm$0.018)~\msun\ from analyses by \citet{Monk1989}
and \citet{Aller1987} of a samples of 9 and 12 LMC and SMC PNe, respectively. Using the {\it Hubble Space Telescope}
to observe a sample of LMC and SMC PNe, \citet{Villaver2003} (LMC) and
\citet{Villaver2004} (SMC) find mean central stars masses of 0.65 and 0.63~\msun, respectively (sample sizes of 16 and 12, respectively).
Adding measurements by \citet{Jacoby1993} and \citet{Dopita1997} to the LMC sample brings the LMC central star mass mean up to 0.67~\msun.
Adding measurements by \citet{Jacoby1993} and \citet{Liu1995} to the SMC sample brings the mean central stars mass up to 0.64~\msun
(the central star of PN J~18 is measured to be 0.63~\msun\ by \citet{Villaver2004}, but 0.56~\msun\ by \citet{Jacoby1993},
once again showing considerable spread). No matter what mean is adopted, and making no
distinction between the LMC and the SMC (whose central star mass distributions are very similar), 
we see that the mean PN central star mass in these two galaxies is larger
than in our Galaxy. 
This is likely due to the fact that both these galaxies are currently star-forming \citep{Gallagher1996,Harris2004,Zaritsky2004},
making the average ages of these two systems younger and their central stars relatively more massive.
From Fig.\ref{CSmassComp.fig} we notice, however, that the minimum central star mass 
encountered in these samples is very similar to the
predicted one.

An obvious difference between our predicted mass distribution and those of observed samples (including that of the WDs,
discussed in \S~\ref{sssec:WDAGB}), is the relatively larger number of more massive central stars in the observed samples.
This might be due to the presence of mergers, which push stars into larger mass bins. This is also the conclusion of
\citet{Liebert2005} regarding the WD sample (\S~\ref{sssec:WDAGB}).

\subsubsection{Birthrates and mass distribution of WDs and the birthrates of AGB stars}
\label{sssec:WDAGB}

\citet{Liebert2005} determined new values for the WD mass distribution and birthrates
from observed samples. The WD mass distribution (Fig.~\ref{CSmassComp.fig}, bottom panel)
has three peaks. They are at 0.565~\msun\ (FWHM=0.188), 0.403~\msun\ 
(FWHM=0.055) and 0.780~\msun\ (FWHM=0.255). The contributions to the three peaks are 76\%, 8\% and
16\%, respectively. \citet{Liebert2005} interpret the low mass peak as those WD that derive from 
post-RGB stars (i.e., that never ascended the AGB),
the main peak as the post-AGB WDs and the high mass peak as those WDs that derive from mergers.
If we take the entire sample of \citet{Liebert2005} of 347 WDs and eliminate the 33 stars with masses smaller than 0.47~\msun\
(these WDs must evolve from stars that {\it never} ascended the AGB since masses this low do not ignite core helium; \citealt{Jimenez1996};
F. Herwig, priv. comm.),
we have a sample of 314 WDs which presumably ascended the AGB. Of these 314 WDs, 242 have masses larger than 0.56~\msun\
and could be
the WDs which are massive enough to have gone through a PN phase. The ratio of the post-PN WDs (242 objects) 
to the total post-AGB
WDs (314 objects) is 0.77. This means that 77\% of all post-AGB WDs have potentially gone though a PN phase, based on core mass
arguments alone (but only 70\% of {\it the entire} WD population has gone through a PN phase). 
This percentage (77\%) compares reasonably well with our ab-initio prediction of ($73 \pm 10$)\%.

The WD {\it recent} birthrates density derived by \citet{Liebert2005} is 
$(1 \pm 0.25) \times 10^{-12}$~WD~yr$^{-1}$~pc$^{-3}$.
Our predicted local (i.e., thin disk) WD formation rate density, ($1.3\pm0.4$)~WD~yr$^{-1} \times$ ($9.0\pm 2.0) \times 10^{-13}$~pc$^{-3}$ = 
($1.2\pm0.4$)$\times 10^{-12}$~WD~yr$^{-1}$~pc$^{-3}$, is close to the value of \citet{Liebert2005}.

It is worth at this point to note that
\citet{Liebert2005} compare their WD
birthrate density to the PN birthrate density of \citet[$2.1\times 10^{-12}$~WD~yr$^{-1}$~pc$^{-3}$]{Phillips2002} and 
of \citet[$3\times 10^{-12}$~WD~yr$^{-1}$~pc$^{-3}$]{Pottasch1996}. The PN/WD birth rate density ratio is thus
2.1--3. \citet{Liebert2005} argue that this ratio should be less then unity
in view of the fact that many WDs did not go through a PN phase (i.e., those that never ascended the AGB as well as
those that never developed a bright PN because of small central star mass leading to slow post-AGB
evolutionary times). They conclude that it
is unclear whether this comparison is telling us something, or whether the PN birthrate densities are too unreliable a number
to be used (the WD birthrate densities is a much more accurate estimate).
Our predicted PN birthrate density (($1.1 \pm 0.5$) $\times$10$^{-12}$~PN~yr$^{-1}$~pc$^{-3}$) being  lower than 
the observationally-based estimates of
\citet{Phillips2002} or \citet{Pottasch1996}, compares more favorably with the WD's.

\citet{Phillips1989} himself addresses the problem outlined above, concluding that 
smaller PN birthrate densities would be more in line with what we know from WDs. However,
he does express the concern that many higher estimates for the PN
birthrate densities are more in line with
other evidence and that lower estimates would not be readily explained.
We point out that if our lower predictions of the PN birthrate density were to be accepted, we would have 
to accept a longer galactic distance scale to derive overall lower local PN densities. This would be even more the case
if we assumed (as we do in Paper~II) that only binaries make PNe, thereby predicting even lower birthrate densities. 

Finally, we turn our attention to the AGB population and seek further constraints of our predictions and
consistency with past results. 90\% (314/357) of the entire WD population has eveolved through the AGB phase and should therefore
be the direct progeny of the AGB population {\it whether or not} it went through a PN phase.
The birthrates of AGB stars should be the same as those of post-AGB WDs and 
larger than those of PNe by the same factor. \citet{Ortiz1996} argue that the AGB star birthrates are between 0.8 and 2.7 AGB stars~yr$^{-1}$, depending on what evolutionary model one adopts. 
Our predicted post-AGB WD birthrate of 2.4 WDs~yr$^{-1}$ (\S~\ref{sec:results})
is within the observed AGB birthrate range, lending further (albeit week) support to our calculation.

\section{The number of PNe in the bulge and in globular clusters}
\label{sec:nrPNbulgeGCs}

Another useful comparison can be carried out with the PN populations of two regions of the Galaxy that
are reasonably well defined: the bulge and the GC system.

\subsection{PNe in the bulge}

\citet{Jacoby2004} predict 250 PNe should reside in the inner 4$\times$4~$\deg$ of the Galaxy. This result is from
their survey, corrected for incompleteness by
assuming a scaled bulge luminosity and using the bright end of the PNLF.
If we assume the bulge within 2~$\deg$ (300~pc) of the center to have a mass of $2-3 \times 10^9$~\msun\  \citep{Lindqvist1992} and using our average-mass model of the bulge ($2.0 \times 10^{10}$~\msun), we can scale the total number of observationally-estimated bulge PNe to the total bulge mass, obtaining $1\,700-2\,500$~PNe from single and stars and binaries. Using the average-mass bulge model, we predict ($7\,200 \pm 3\,000$) PNe, which is 3--5
times the amount estimated from observations.
The over-abundance of predicted vs. ``observed" PNe is in line
with the comparison carried out for the entire Galaxy. 

\subsection{PNe in globular clusters}

For the GCs the situation is more complicated.
The number of PNe contributed by the spheroid can be used to predict the number of PNe in GCs,
assuming they have similar star formation histories.  
There are about 150 GCs with an average of 300\,000 stars each\footnote{This number is an extrapolation
from GC luminosities and luminosity functions (data from \citet{Harris1996}). It is likely to be accurate to within a factor of 2-3;
see also later.}.
We assume that all GCs have the same mean age of 11.5~Gyr \citep{Rakos2005}, which also corresponds to 
the peak period of star formation in the spheroid (since the spheroid has an average metallicity of
$\log(Z/Z_{\odot})=-1.5$ and an average turn-off mass of 0.85~\msun).  

Using any of the three IMFs from \S~\ref{ssec:IMF},
the average mass of a star in a GC is $\approx$0.45~\msun. This means that the total mass of the galactic GC system is
$2.0\times10^7$~\msun\ 
($0.45\times300\,000\times150$; cf. this number with the value $2\times10^7$~\msun\ of \citet{Pena1995}). 
Scaling the number of spheroid PNe (70$\pm$80) by the ratio of the spheroid and GC system masses, the predicted
number of PNe in GCs is 0--1.5 PNe
( ($70\pm80$)$\times$(2.0 $\times 10^7$) / (2$\times 10^9$);
where $2\times10^9$~\msun\ is the mass of the spheroid; \S~\ref{sssec:massSpheroid}). The main reason for the large 
uncertainty, is the uncertainty in the mass cut-off value for PN production (\S~\ref{ssec:PNLifetimes}), which primarily
affects old populations.  
In fact, when the cutoff mass is taken to be 0.95~\msun\ (the largest of the three values we use in
\S~\ref{ssec:PNLifetimes}), 
no GC PN is produced in any of the models.  

The total number of PNe found in the galactic GC system is four \citep{Jacoby1997}. 
\citet{Jacoby1997} determined that, by applying the fuel consumption method of \citet{Renzini1986}, 
16 PNe are expected to reside in the Milky Way GC system. However, they conceded that this 
method does not account for the fact that low mass
central stars evolve too slowly to ionize the PN before the circumstellar gas is dispersed (\S~\ref{ssec:PNLifetimes}).
If a central star cut-off mass is assumed, no PN might be expected to reside in GCs at all, 
in line with our prediction and at odds with the observations.

Looking at the characteristics of the four GC PNe we find that two of them have a very high central
star mass (K648 has a derived central star mass of 0.62~\msun, while IRAS~18333-2357 has a central stars
mass of 0.75~\msun; \citealt{Alves2000}; \citealt{Harrington1993}). 
These masses correspond to main sequence masses (2.3 and 3.6~\msun, respectively; \citealt{Weidemann2000}),
that are up to three times the cluster turn-off mass,
implying some kind of merger formed the progenitors of these central stars. This could be a common envelope merger
or a merger that occurred during the main sequence and that temporarily generated a blue straggler star, which later
evolved into a PN (but see \citet{Ciardullo2005} for a connection between bright PNe and blue stragglers).
One of the four GC PNe (IRAS!18333-2357) is hydrogen-deficient, a characteristics displayed by only 5 galactic PNe
\citep{Harrington1996}, making this a very rare phenomenon. Hydrogen-deficient PNe are attributed 
to the action of a last helium thermal 
pulse that ejects and ingests the remaining hydrogen envelope of a central star \citep{Iben1983, Herwig2000}. 
However, others (e.g., \citealt{DeMarco2002})
have argued that binary mergers might be responsible for these objects. While the jury is still out on hydrogen-deficient
PNe, we might speculate that the high incidence of H-deficient PNe in the GC population (although
we are dealing with very low number statistics), is in line with the idea that these objects
derive from merged stars.

One could therefore argue that it is possible that none of the observed GC PNe is a ``regular" PN and that all of them 
are nebulosities deriving from some kind of binary interaction, favored in the crowded environments of GCs, because of the high frequency of orbit-reducing stellar exchanges \citep{Hurley2003}.
This possibility is in line with the finding by \citet{Jacoby1997} that
the presence of a PN in a GC
appears to correlate with the number of X-ray sources, which are in turn associated to low mass X-ray binaries.

The issue of the presence of PN in GCs is not a simple one and cannot be used as a model constraint. On the other hand, if indeed GC PNe are merger products, the well determined distances to GCs 
will make these four PNe ideal for case studies of this elusive class of objects.

\section{Conclusion}
\label{sec:conclusion}

In this paper we have calculated the number of PNe expected to reside in the Galaxy, if PNe derive from single stars as well
as from stars in binary systems. The resulting value, $(4.6 \pm 1.3) \times 10^4$
objects with radius smaller than 0.9~pc, is six times higher than the observationally-based estimate of 7200$\pm$1800 objects with the same radius constraints (discrepant at the 2.9$\sigma$ level). Although it could be argued that the actual galactic PN population is larger, we stress that our predicted value is not absolute (since it is constrained by a maximum PNe radius), but is rather designed to be compared to the determination of \citet{Peimbert1993} and as such the comparison is much more reliable than the individual figures.

We argue that the discrepancy is due to the fact that not all stars otherwise fit to make a PN, actually do make a PN. 
In Paper~II we will calculate the galactic PN population deriving from binary interactions and 
show that, based on population synthesis arguments alone, it is more likely that most PNe could derive from such interaction rather than being produced by single stars.

We also compared 
our predicted PN birthrate density with that derived observationally from counts of local PNe.
Our prediction (($1.1\pm0.5)\times$10$^{-12}$~PN~yr$^{-1}$~pc$^{-3}$) is lower than most observationally-based estimates
(2.1$\times$10$^{-12}$~PN~yr$^{-1}$~pc$^{-3}$; \citealt{Phillips2002})\footnote{It will strike the reader that from observational PN birthrate density
values that are generally {\it larger} than our prediction, total galactic PN populations are derived 
that are {\it lower} than our estimate (see for instance \citet{Phillips2002}).
This is due to the very different methods to derive the birthrate densities in
our theoretical derivation and in the observationally-based derivations found in
the literature.}, although it is within the quoted uncertainty. Our estimate is similar to the more reliable observational estimate of the WD birthrate density (1.0$\times$10$^{-12}$~WD~yr$^{-1}$~pc$^{-3}$; \citet{Liebert2005}).
The PN birthrate density from binary interaction we will derive in Paper~II 
will clearly be lower. A discussion of the implication of this discrepancy 
is left to that paper.

We predict that only 73\% of post-AGB WDs went through the PN phase. 
This is in line
with what we know of post-AGB evolutionary timescales (namely, that central stars with mass $<$0.56~\msun\ do not make a PN), 
the observed mass distributions of central stars (minimum central stars mass $\sim$0.56~\msun)
and the mass distribution of WDs (which indicates that only 77\% of post-AGB WDs have masses large enough to go through a PN phase.). 

\acknowledgments

We are grateful for useful conversations with Falk Herwig, George Jacoby and 
Mordecai-Mark Mac Low. We are also indebted to several of the participants of the IAU Symposium 234 on planetary nebulae, who have given us comments on several aspects of this paper. In particular we are grateful for the comment of David Frew and Manuel Peimbert. We also than the referee,
J.P. Phillips, for an excellent review of this manuscript that helped improve the paper as a whole. MM gratefully acknowledges the 
Research Experience for Undergraduate program of the National Science Foundation (REU grant \# AST~0243837).

\bibliographystyle{apj}                       %% AASTeX
\bibliography{bibliography}

\clearpage

\clearpage

\clearpage

\begin{figure}
\centering
\includegraphics[width=5.5in]{fig14.eps}
\caption{Histograms of central stars masses. From top to bottom we compare: our population synthesis results
(see also Fig.~\ref{pnfinalmassdist.fig}), the galactic sample of \citet{Gorny1997} (photometric method), 
the galactic sample of \citet{Rauch1998}, \citet{Rauch1999} and \citet{Napiwotzki1999} (spectroscopic method), the
combined LMC and SMC samples of 
\citet{Villaver2003}, 
\citet{Villaver2004}, 
\citet{Jacoby1993}, \citet{Dopita1997},
and \citet{Liu1995} 
and, finally, the WD sample of \citet{Liebert2005}. The vertical line marks central star
mass 0.56~\msun.}
\label{CSmassComp.fig}
\end{figure}

\begin{deluxetable}{lccc}
%\tabletypesize{\scriptsize}
\tablecaption{Mass estimates for the four galactic components.\label{tab:mass}}
\tablehead{\colhead{Component} & \multicolumn{3}{c}{Masses ($10^9$~\msun)} }
\startdata
&low &medium &high \\
Thin Disk& 29 & 36 & 43 \\
Thick Disk& 4 & 4 & 4 \\
Bulge & 16 & 20 & 24 \\
Spheroid & 2 & 2 & 2 \\
Total & 51 & 62 & 73 \\
\enddata
\end{deluxetable}

\begin{deluxetable}{lccccc}
%\tabletypesize{\scriptsize}
\tablecaption{Parameters used in the star formation rate equations , \ref{exp.eqn} and \ref{sandage.eqn}.\label{tab:SFR}}
\tablehead{\colhead{Component}&{$t_0$}&{$t_{end}$}&{Mean Age}&{$\tau$(Eq.\ref{exp.eqn})}&{$\tau$(Eq.\ref{sandage.eqn}) }}
\startdata
Spheroid & 0.0 & 4.5 & 11.5 & 2.09 & 1.25 \\
Thick Disc & 1.0 & 6.0 & 10.0 & 4.06 & 1.62 \\
Bulge - Young & 0.5 & 13.0 & 9.0 & 4.14 & 2.79 \\
Bulge - Average & 0.5 & 13.0 & 10.0 & 2.60 & 1.99 \\
Bulge - Old & 0.5 & 13.0 & 11.0 & 1.50 & 1.20 \\
Thin Disc - Young & 3.0 & 13.0 & 4.5 & -16.6 & 5.33 \\
Thin Disc - Average & 3.0 & 13.0 & 5.5 & 16.6 & 3.74 \\
Thin Disc - Old & 3.0 & 13.0 & 6.5 & 5.25 & 2.80 \\

\enddata
\end{deluxetable}

\begin{deluxetable}{lccc}
%\tabletypesize{\scriptsize}
\tablecaption{Percentage errors incurred from each step. \label{tab:Error}}
\tablehead{\colhead{Parameter}&{\% Error}&{$\sigma_{\% Error}$}}
\startdata

SFH Age & 22\% & 4\% \\
Mass & 19\% & 0.2\% \\
IMF & 17\% & 0.5\% \\
Cutoff Mass & 9\% & 4\% \\
SFH Type & 5\% & 2\% \\
PN Visibility & 1\% & 1\% \\

\enddata
\end{deluxetable}
\end{document}